\documentclass[10pt]{article}
\usepackage{latexsym,amssymb,amsfonts,amsmath}
\usepackage{graphicx}

\newcommand{\sech}{ {\rm sech} }

\begin{document}
\title{The multiple soliton and peakon solutions of the Dullin-Gottwald-Holm equation}
 \author{{Zhaqilao${{\thanks {
 Corresponding author: Tel. : +86 471 4392483 : fax: +86 471 7383390. E-mail address:
zhaqilao@imnu.edu.cn}}}$}
\\
{\scriptsize{College of Mathematics Science, Inner Mongolia Normal
University, Huhhot 010022, People's Republic of China}}}

\date{}
\maketitle

%E-mails: zhaqilao@imnu.edu.cn and 1416528264@qq.com (Yuanli  Li)

\baselineskip 18pt  {\bf \large{Abstract:}}  Explicit multi-soliton and multi-peakon solutions of the Dullin-Gottwald-Holm equation are constructed via Darboux transformation and direct computation, respectively. To this end we first map the Dullin-Gottwald-Holm equation to a negative order KdV equation by a reciprocal transformation. Then we use the Darboux matrix approach to derive multi-soliton solutions of the Dullin-Gottwald-Holm equation from the solutions of the negative order KdV equation. Finally, we find multi-peakon solutions of the Dullin-Gottwald-Holm equation in weak sense. For $\alpha=\frac{1}{2}$ and $\alpha\neq \frac{1}{2}$, several types of two-peakon solutions are discussed in detail. Moreover, the dynamic behaviors of the obtained solutions are illustrated through some figures.

\vskip 1mm



%\noindent \underline{\hspace*{14.8cm}}

\parskip 10pt
\setcounter{equation}{0}
\section{Introduction}
In 2001, Dullin, Gottwald, and Holm discussed the following $1+1$ quadratically nonlinear equation (DGH) \cite{1} for a unidirectional water waves with fluid velocity $u(x,t)$,
\begin{equation}\label{1}
m_t+c_0 u_x+u m_x+2m u_x=-\gamma u_{xxx}.
\end{equation}
Here $m=u-\alpha^2 u_{xx}$ is a momentum variable, partial derivatives are denoted by subscripts, the constants $\alpha^2$ and $\gamma/c_0$ are squares of length scales, and $c_0=\sqrt{gh}$ is the critical shallow water speed for undisturbed water at rest at spatial infinity, where $h$ is the mean fluid depth and $g$ is the gravitational constant $g=9.8m/s^2$. The DGH equation is related to two separately integrable soliton equations for water waves, when $\alpha=0$ and $\gamma\neq 0$, Eq. (\ref{1}) becomes the Korteweg-de Vries (KdV) equation \cite{2}
\begin{equation}\label{200}
u_t+c_0 u_x+3u u_x+\gamma u_{xxx}=0.
\end{equation}
And, setting $\gamma=0$ and $\alpha=1$, Eq. (\ref{1}) implies the Camassa-Holm (CH) equation \cite{3}
\begin{equation}\label{201}
u_t+c_0 u_x-u_{xxt}+3uu_x=2u_xu_{xx}+uu_{xxx}.
\end{equation}
Hence, Eq. (\ref{1}) is an important integrable shallow water wave equation. It has a soliton-liking KdV equation and peakon-liking CH equation. Dullin et al \cite{1} showed that Eq. (\ref{1}) has a bi-Hamiltonian structure and a Lax pair formulation.

Indeed, the DGH equation  (\ref{1}) has been studied in a lot of works  \cite{4}-\cite{17}. Guo and Liu \cite{4,5} obtained expressions of cusp wave and peaked wave solutions. Tian et al \cite{6}, discussed the Cauchy problem and the scattering problem of Eq. (\ref{1}). Ai and Gui \cite{7} presented an algorithm for the inverse scattering problem and studied the low regularity solution of Eq. (\ref{1}). In \cite{8,9}, the orbital stability of one single solitary waves and the sum of $N$ peakons for Eq. (\ref{1}) have been proved. In \cite{10}-\cite{15}, some peakons, kinks, compactons, solitons, solitary patterns solutions and exact solutions are obtained.

In the case $\gamma=-c_0\alpha^2$, the equation (\ref{1}) becomes
\begin{equation}\label{2}
m_t+c_0 m_x+um_x+2 mu_x=0,\,\,\,m=u-\alpha^2 u_{xx},
\end{equation}
where $c_0$ is a constant. The existence and uniqueness of global weak solutions to the DGH equation (\ref{2}) provided the initial data satisfies certain conditions were derived in \cite{16}.  It was shown in \cite{1} that the DGH equation (\ref{2}) possesses the following Lax pair with the spatial part
\begin{equation}\label{3}
\psi_{xx}=\dfrac{1}{4}(\dfrac{1}{\alpha^2}+4\lambda m)\psi,
\end{equation}
and the temporal part
\begin{equation}\label{4}
\psi_{t}=(\dfrac{1}{2\alpha^2 \lambda}-c_0-u)\psi_x+\dfrac{1}{2}u_x\psi,
\end{equation}
where $\lambda$ is a spectral parameter.

Chen et al \cite{17} obtained multiple soliton solutions of Eq. (\ref{2}) with the Darboux transformation method. Motivated by the interest in the approaches in \cite{17}-\cite{18} and \cite{19}-\cite{21}, we study the multi-soliton, multi-peakon solutions to the DGH equation (\ref{2}) and their interaction by using the numerical simulation methods. The paper is organized as follows. In Section 2, the multi-soliton solutions of the DGH equation (\ref{2}) are constructed by Darboux matrix approach. In Section 3, the several types of two-peakon solutions of the DGH equation (\ref{2}) are derived by different integrable constants. The conclusions are given in Section 4.

\section{Multi-soliton solution}
The Liouville transformation \cite{22}
\begin{equation}\label{5}
r=\sqrt{m},
\end{equation}
converts the DGH equation into conservation form
\begin{equation}\label{6}
r_t+[(c_0+u)r]_x=0.
\end{equation}
It permits us to define a reciprocal transformation $(x,t)\mapsto (y,\tau)$ by
\begin{equation}\label{7}
dy=rdx-(c_0+u)r dt,\,\,\,d\tau=dt,
\end{equation}
and we have
\begin{equation}\label{8}
\dfrac{\partial}{\partial x}=r\dfrac{\partial}{\partial y},\,\,\,
\dfrac{\partial}{\partial t}=\dfrac{\partial}{\partial \tau}-(c_0+u)r \dfrac{\partial}{\partial y}.
\end{equation}

With the help of the reciprocal transformation (\ref{7}), the DGH equation (\ref{2}) is transformed to
\begin{equation}\label{9}
r_\tau+r^2u_y=0,
\end{equation}
\begin{equation}\label{10}
u=r^2-\alpha^2 r(\ln r)_{y\tau}.
\end{equation}
According above role we set $\psi(x,t)=r^{-\frac{1}{2}}(y,\tau) \phi(y,\tau)$, the Lax pair (\ref{3}) and (\ref{4}) become
\begin{equation}\label{11}
\phi_{yy}=\lambda \phi+Q \phi,
\end{equation}
\begin{equation}\label{12}
\phi_{\tau}=\dfrac{1}{4\alpha^2 \lambda}(2 r \phi_y-r_y \phi),
\end{equation}
where $Q=\dfrac{r_{yy}}{2r}-\dfrac{r^2_y}{4r^2}+\dfrac{1}{4\alpha^2 r^2}$.

From the compatibility condition of the Lax pair (\ref{11}) and (\ref{12}), we derive a soliton integrable equation
\begin{equation}\label{13}
\alpha^2 Q_\tau=r_y,\,\,\,\,\,r_{yyy}-4Qr_y-2Q_yr=0,
\end{equation}
which is the negative order KdV (NKdV) equation \cite{23}.

In order to study the Darboux transformation (DT) for the Lax pair (\ref{11}) and (\ref{12}), first we rewrite the equation (\ref{11}) and (\ref{12}) in the following matrix form
\begin{equation}\label{14}
\Phi_y=U\Phi=\left(\begin{array}{cc}  0 &  1\\[1mm]
\lambda+Q  & 0
\end{array}
\right)\left(\begin{array}{cc}  \phi\\[1mm]
\phi_{y}
\end{array}
\right),
\end{equation}
\begin{equation}\label{15}
\Phi_\tau=V\Phi=\left(\begin{array}{cc}  -\dfrac{r_y}{4\alpha^2 \lambda} &  \dfrac{r}{2 \alpha^2 \lambda}\\[3mm]
\dfrac{r}{2\alpha^2}+\dfrac{Qr}{2\alpha^2 \lambda}-\dfrac{r_{yy}}{4\alpha^2 \lambda}  & \dfrac{r_y}{4\alpha^2 \lambda}
\end{array}
\right)\left(\begin{array}{cc}  \phi\\[3mm]
\phi_{y}
\end{array}
\right).
\end{equation}

In fact, one just checks that the compatibility condition $\Phi_{y\tau}=\Phi_{\tau y}$, namely, $U_\tau-V_y+[U,V]=0$ generates the equations (\ref{13}).

We now introduce a transformation
\begin{equation}\label{16}
\overline{\Phi}=T\Phi,
\end{equation}
which transform the Lax pair (\ref{14}) and (\ref{15}) into a Lax pair of $\overline{\Phi}$
\begin{equation}\label{17}
\overline{\Phi}_y=\overline{U}\,\overline{\Phi},\,\,\,\overline{U}=(T_y+TU)T^{-1},
\end{equation}
\begin{equation}\label{18}
\overline{\Phi}_\tau=\overline{V}\,\overline{\Phi},\,\,\,\overline{V}=(T_\tau+TV)T^{-1},
\end{equation}
where $\overline{U}$ and $\overline{V}$ have the same as $U$ and $V$ except that $Q$ and $r$ are replaced by $\overline{Q}$ and $\overline{r}$, respectively. It is clear that $(\overline{Q},\,\,\overline{r})$ is a new solution of (\ref{13}).

According to the form of $U$, we select the elementary Darboux matrix of the KdV hierarchy
\begin{equation}\label{19}
T=\left(\begin{array}{cc}  -\delta &  1   \\[3mm]
\delta^2-\lambda_1+\lambda  & -\delta
\end{array}
\right),\,\,\,\,\det T=\lambda_1-\lambda,
\end{equation}
where
\begin{equation}\label{20}
\delta=\dfrac{\phi_{1y}}{\phi_1},
\end{equation}
and $\phi=(\phi_1,\phi_{1y})^T$ is a solution of the Lax pair (\ref{14}) and (\ref{15}) with $\lambda=\lambda_1$. The function $\delta$ satisfies the following Riccati equations
\begin{equation}\label{21}
\delta_y=\lambda_1+Q-\delta^2,
\end{equation}
\begin{equation}\label{22}
\delta_\tau=\dfrac{r}{2\alpha^2}+\dfrac{Qr}{2\alpha^2 \lambda_1}-\dfrac{r_{yy}}{4\alpha^2 \lambda_1}
+\dfrac{r_y}{2 \alpha^2 \lambda_1}\delta-\dfrac{r}{2\alpha^2 \lambda_1}\delta^2.
\end{equation}

From the above discussions, we give the following propositions.

\textbf{Proposition 1.} The matrix $\overline{U}$ determined by the second expression of (\ref{17}) has the same form as $U$, namely
\begin{equation}\label{23}
\overline{U}=\left(\begin{array}{cc}  0 &  1\\[1mm]
\lambda+\overline{Q}  & 0
\end{array}
\right),
\end{equation}
where the transformation formula from original potential $Q$ into new one is given by
\begin{equation}\label{24}
\overline{Q}=Q-2\delta_y=-2\lambda_1-Q+2\delta^2.
\end{equation}

\textbf{Proof.}  Substituting $T$ and $U$ into the right hand side of the second expression of (\ref{17}) and (\ref{21}), we can get the conclusion
\begin{equation}\label{25}
(T_y+TU)T^{-1}=\left(\begin{array}{cc}  0 &  1   \\[3mm]
\lambda+(-2\lambda_1-Q-2\delta^2)         &  0
\end{array}
\right)=
\left(\begin{array}{cc}  0 &  1   \\[3mm]
\lambda+\overline{Q}       &  0
\end{array}
\right)=\overline{U}.
\end{equation}
The proof is completed.

\textbf{Proposition 2.} The matrix $\overline{V}$ determined by the second expression of (\ref{18}) has the same form as $V$, in which original potentials $Q$ and $r$ mapped into new potentials $\overline{Q}$ and $\overline{r}$ according to the transformation
\begin{equation}\label{26}
\begin{array}{ll}
\overline{Q}=Q-2\delta_y=-2\lambda_1-Q+2\delta^2,\\[2mm]
\overline{r}=r-2\alpha^2\delta_\tau=\dfrac{1}{2\lambda_1}(-2Qr+2r\delta^2-2r_y\delta+r_{yy}).
\end{array}
\end{equation}

With the aid of (\ref{21}) and (\ref{13}), we list $r_y$ and $r_{yy}$ as follows
\begin{equation}\label{27}
\overline{r}_y=-\dfrac{1}{\lambda_1}(-2r\delta(\lambda_1+Q-\delta^2)+(\lambda_1-2\delta^2)r_y+\delta r_{yy}),
\end{equation}
\begin{equation}\label{28}
\begin{array}{ll}
\overline{r}_{yy}=\dfrac{1}{\lambda_1}(2r(\lambda_1^2+Q^2-4\lambda_1\delta^2+3\delta^4+2Q(\lambda_1-2\delta^2))\\[3mm]
\mbox{} \hskip 1.0cm+2(3\lambda_1+Q)\delta r_y-6\delta^3 r_y-(2\lambda_1+Q)r_{yy}+3\delta^2 r_{yy}).
\end{array}
\end{equation}

\textbf{Proof.} Substituting $T$ and $V$ into the right hand side of the second expression of (\ref{18}) and using (\ref{27})-(\ref{28}), we can get the conclusion
\begin{equation}\label{29}
(T_\tau+TV)T^{-1}=\left(\begin{array}{cc}  \dfrac{1}{\lambda}q^{(-1)}_{11} &  \dfrac{1}{\lambda}q^{(-1)}_{12}   \\[3mm]
q^{(0)}_{21}+\dfrac{1}{\lambda}q^{(-1)}_{21}         &  -\dfrac{1}{\lambda}q^{(-1)}_{11}
\end{array}
\right):=\Gamma,
\end{equation}
where
\begin{equation}\label{30}
\begin{array}{ll}
q^{(-1)}_{12}=q^{(0)}_{21}=\dfrac{1}{2\alpha^2\lambda_1}(-Qr+r\delta^2-\delta r_y+\dfrac{1}{2}r_{yy})=\dfrac{1}{2\alpha^2}\overline{r},\\[3mm]
q^{(-1)}_{11}=-\dfrac{1}{4\alpha^2}(-\dfrac{1}{\lambda_1}(-2r\delta(\lambda_1+Q-\delta^2)+(\lambda_1-2\delta^2)r_y+\delta r_{yy}))=-\dfrac{1}{4\alpha^2}\overline{r}_y,\\[3mm]
q^{(-1)}_{21}=\dfrac{1}{4\alpha^2\lambda_1}(-2\lambda_1-Q+2\delta^2)(-2Qr+2r\delta^2-2\delta r_y+r_{yy})\\[3mm]
\mbox{} \hskip 1.0cm-\dfrac{1}{4\alpha^2}[\dfrac{1}{\lambda_1}(2r(\lambda_1^2+Q^2-4\lambda_1\delta^2+3\delta^4+2Q(\lambda_1-2\delta^2))\\[3mm]
\mbox{} \hskip 1.0cm+2(3\lambda_1+Q)\delta r_y-6\delta^3 r_y-(2\lambda_1+Q)r_{yy}+3\delta^2r_{yy})]\\[3mm]
\mbox{} \hskip 1.0cm+\dfrac{1}{2\alpha^2}[\dfrac{1}{2\lambda_1}(-2Qr+2r\delta^2-2r_y\delta+r_{yy})]\\[3mm]
\mbox{} \hskip 1.0cm
=\dfrac{\overline{Q}\,\,\overline{r}}{2\alpha^2}-\dfrac{\overline{r}_{yy}}{4\alpha^2}+\dfrac{\overline{r}}{2\alpha^2}.
\end{array}
\end{equation}
Thus $\Gamma=\overline{V}$. This completes the proof.

According to propositions 1 and 2, the transformations (\ref{26}) map the Lax pair (\ref{14})-(\ref{15}) into the Lax pair
 (\ref{17})-(\ref{18}) with the same type. Therefore, both of the Lax pairs lead to the same soliton equation (\ref{13}). The transformations (\ref{26}) is usually called DT of the soliton equation (\ref{13}). With the help of the DT (\ref{26}) and proposition 1 and 2, the mew solutions of the soliton equation (\ref{13}) can be generated with the seed solution (any constant value $(Q_0,r_0)$ is a solution).

 To complete the solution $u(x,t)$ of the DGH equation (\ref{2}), we have only to obtain the coordinate transformation between $x$, $t$ and $y$, $\tau$. When $\lambda=0$, equations (\ref{3}) and  (\ref{11}) become
\begin{equation}\label{100}
\psi_{xx}=\dfrac{1}{4\alpha^2}\psi,
\end{equation}
\begin{equation}\label{101}
\phi_{yy}=Q\phi.
\end{equation}
For equation (\ref{100}), we may choose a solution $\psi$ to be $\psi=e^{\frac{x}{2\alpha}}$. For equation (\ref{101}), we have a solution $\phi^{+}_0=e^{\sqrt{Q_0}y}$ (or $\phi^{-}_0=e^{-\sqrt{Q_0}y}$) and another solution can be obtained by
\begin{equation}\label{102}
\left(\begin{array}{cc}  \phi_{[1]}   \\[3mm]
\phi_{[1]y}
\end{array}
\right)=T|_{\lambda=0}\left(\begin{array}{cc}  \phi^{+}_0   \\[3mm]
\phi^{+}_{0y}
\end{array}
\right)=\left(\begin{array}{cc}  -\delta & 1  \\[3mm]
\delta^2-\lambda_1   &  -\delta
\end{array}
\right)\left(\begin{array}{cc}  \phi^{+}_0   \\[3mm]
\phi^{+}_{0y}
\end{array}
\right).
\end{equation}
Thus the relation $\phi_{[1]}=r^{\frac{1}{2}}\psi$ yields
\begin{equation}\label{103}
e^{\frac{x}{2\alpha}}=\dfrac{\phi_{[1]}}{\sqrt{r}},
\end{equation}
which gives rise to the following relation
\begin{equation}\label{104}
x=\alpha \ln \dfrac{\phi^2_{[1]}}{r}=\alpha \ln \dfrac{(-\delta \phi^{+}_0+\phi^{+}_{0y})^2}{r}.
\end{equation}
On the other hand, from (\ref{7}) it is easy to see  $t=\tau$.

If starting from a nonzero seed solution $(Q_0,r_0)$ of the equation (\ref{13}) and substituting this seed solution into the Lax pair (\ref{14})-(\ref{15}), their two basic solutions are chosen as

\begin{equation}\label{31}
\Phi=\left(\begin{array}{cc}  \phi_j   \\[3mm]
\phi_{jy}
\end{array}
\right)=\left(\begin{array}{cc}  \cosh\xi_j   \\[3mm]
\Omega_j\sinh\xi_j
\end{array}
\right),\,\,\,\,\textmd{j is odd number},
\end{equation}
and
\begin{equation}\label{32}
\Phi=\left(\begin{array}{cc}  \phi_j   \\[3mm]
\phi_{jy}
\end{array}
\right)=\left(\begin{array}{cc}  \sinh\xi_j   \\[3mm]
\Omega_j\cosh\xi_j
\end{array}
\right),\,\,\,\,\textmd{j is even number},
\end{equation}
where
\begin{equation}\label{33}
\xi_j=\Omega_j\left(y+\dfrac{r_0}{2\alpha^2 \lambda_j}\tau\right),\,\,\,\Omega_j=\sqrt{Q_0+\lambda_j}.
\end{equation}

By applying (\ref{31})-(\ref{32}), (\ref{26}) and (\ref{10}), we can obtain $N$-soliton solutions for the DGH equation (\ref{2}).

\noindent \textbf{Case 1} ($j=1$). Noticing equation (\ref{20}), we have
\begin{equation}\label{34}
\delta_1=\Omega_1\tanh\xi_1,
\end{equation}
where $\xi_1=\Omega_1\left(y+\dfrac{r_0}{2\alpha^2 \lambda_1}\tau\right)$ and $\Omega_1=\sqrt{Q_0+\lambda_1}$.

Substituting (\ref{34}) into (\ref{26}), we obtain
\begin{equation}\label{35}
Q_1=Q_0-2\Omega^2_1\sech^2\xi_1,\,\,\,r_1=r_0-\dfrac{r_0\Omega^2_1}{\lambda_1}\sech^2\xi_1.
\end{equation}
Substituting $r_1$ and $\delta_1$ into (\ref{10}) and (\ref{104}), we arrive at the one-soliton solution of the DGH equation (\ref{2})
\begin{equation}\label{36}
\begin{array}{ll}
u=r^2_1-\alpha^2r_1(\ln r_1)_{y\tau}=\dfrac{(\lambda^2_1\sinh^2\xi_1+\lambda_1Q_0+2Q_0^2)r^2_0}{\lambda_1(\lambda_1\sinh^2\xi_1-Q_0)},\\[5mm]
x=\alpha \ln \dfrac{(-\delta_1 \phi^{+}_0+\phi^{+}_{0y})^2}{r_1}
=\alpha \ln\dfrac{e^{2\sqrt{Q_0}y}(\sqrt{Q_0}-\Omega_1\tanh\xi_1)^2}
{r_0-\dfrac{r_0\Omega^2_1}{\lambda_1}\sech^2\xi_1},\\[3mm]
t=\tau.
\end{array}
\end{equation}
This parametric representation of the one-soliton solution of the DGH equation (\ref{2}) is illustrated in figure 1, which is a right-going travelling wave. In figure 2, we have adjusted the values of $\alpha$ ( as $\alpha=1,2,3$) so that to have solitary waves shown there, if $\alpha$ has greater value, then wave has wider body and slower speed.
\begin{center}
\resizebox{2.7in}{!}{\includegraphics{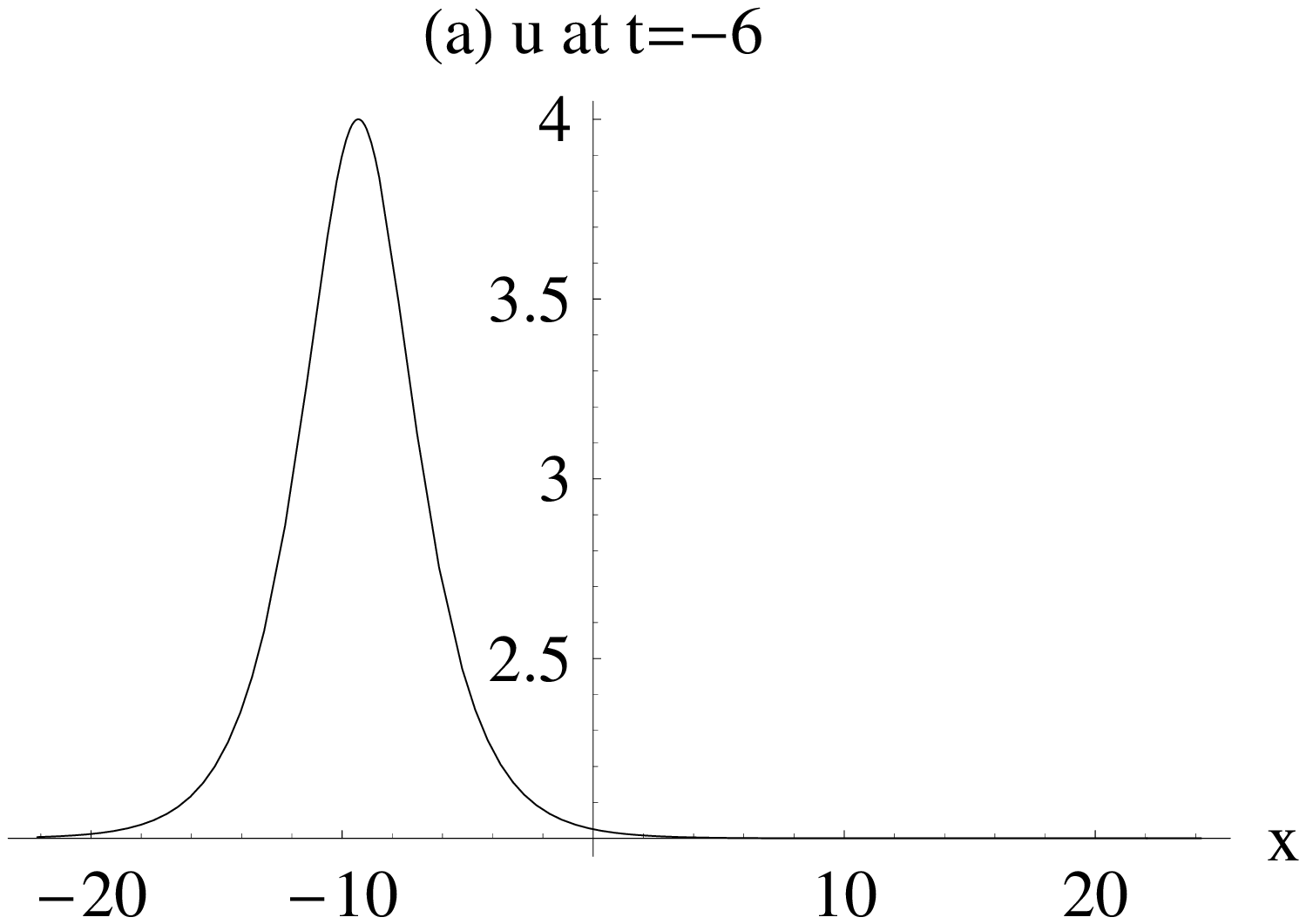}}\,\,\,
\resizebox{2.7in}{!}{\includegraphics{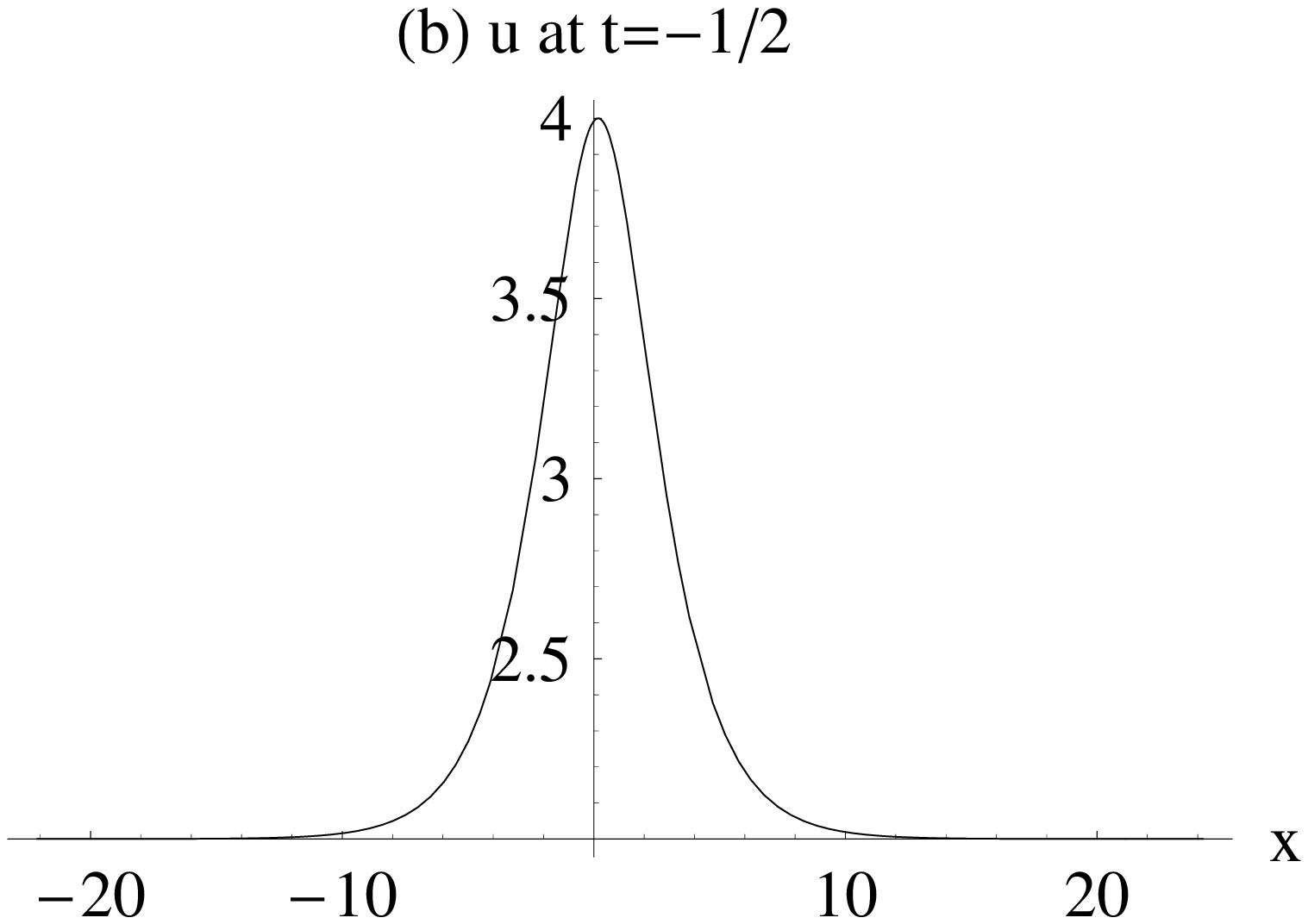}}\,\,\,
\resizebox{2.7in}{!}{\includegraphics{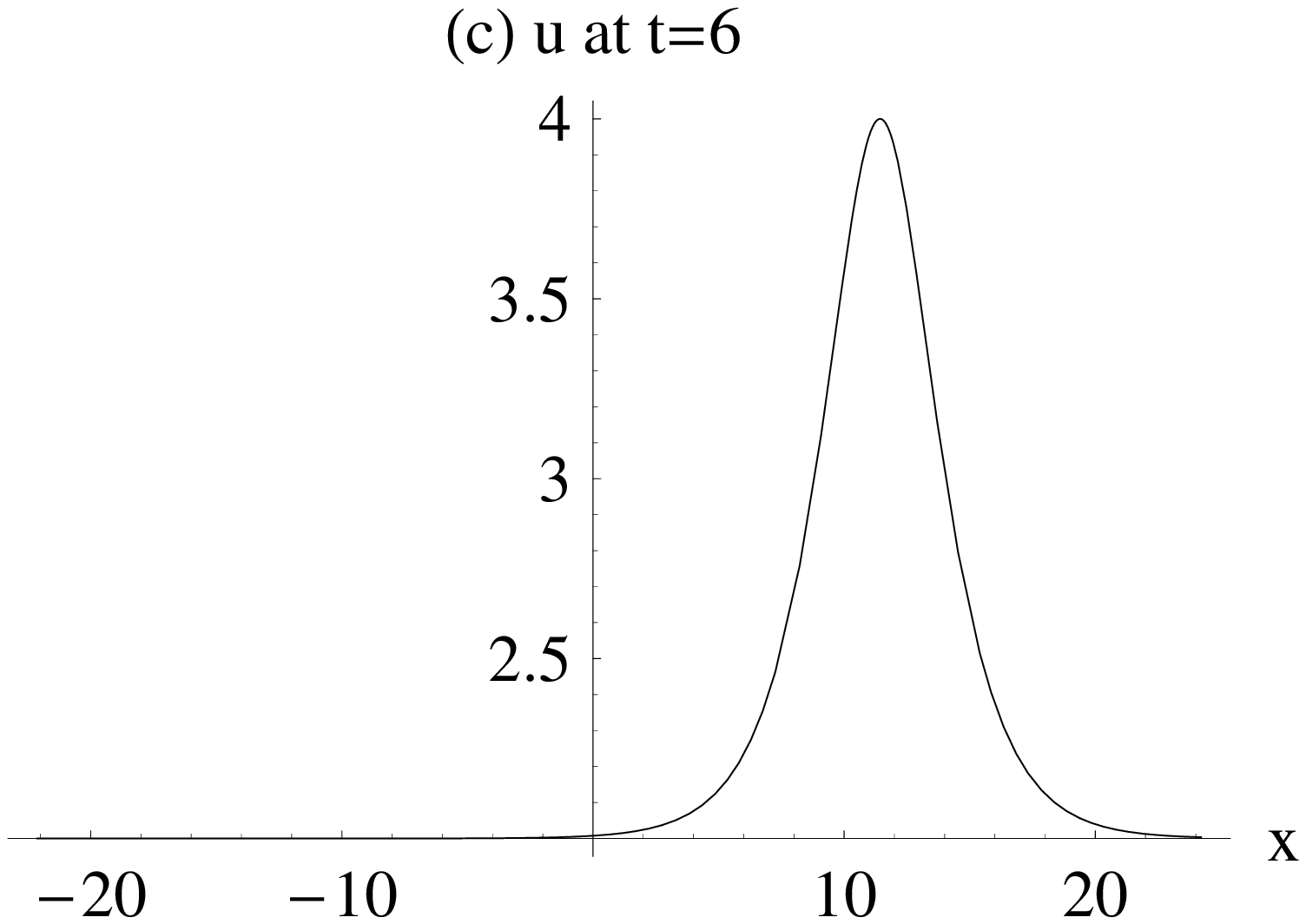}}
\end{center}
\centerline{\small{Figure 1.\,\,One-soliton solution defined by (\ref{36}) with $Q_0=3/2$, $r_0=\sqrt{2}$, $\lambda_1=-1$, $\alpha=1$,}}
\centerline{\small{and (a) $u$ at $t=-6$,  (b) $u$  at $t=-1/2$, (c) $u$  at $t=6$.}}

\begin{center}
\resizebox{2.7in}{!}{\includegraphics{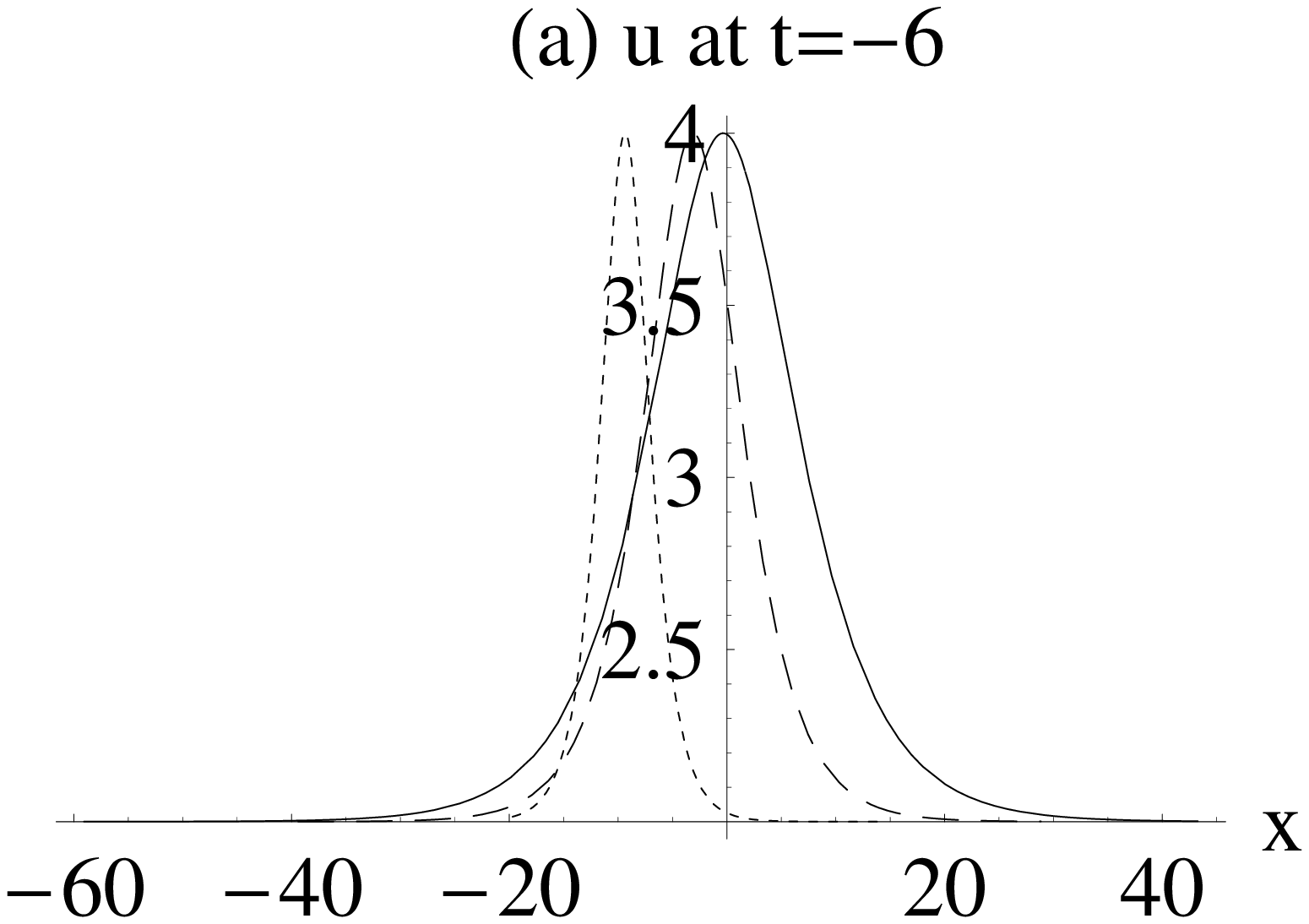}}\,\,\,
\resizebox{2.7in}{!}{\includegraphics{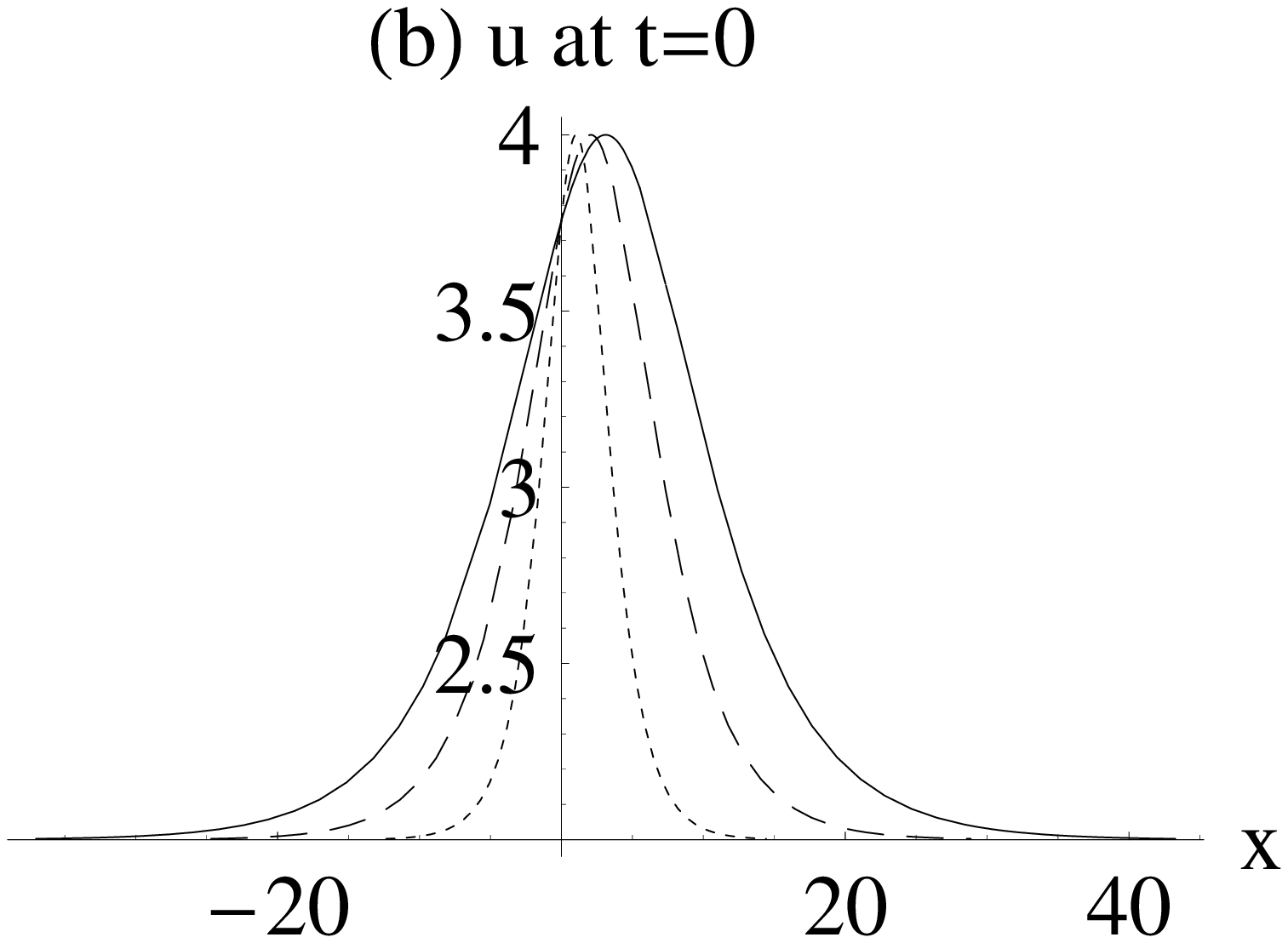}}\,\,\,
\resizebox{2.7in}{!}{\includegraphics{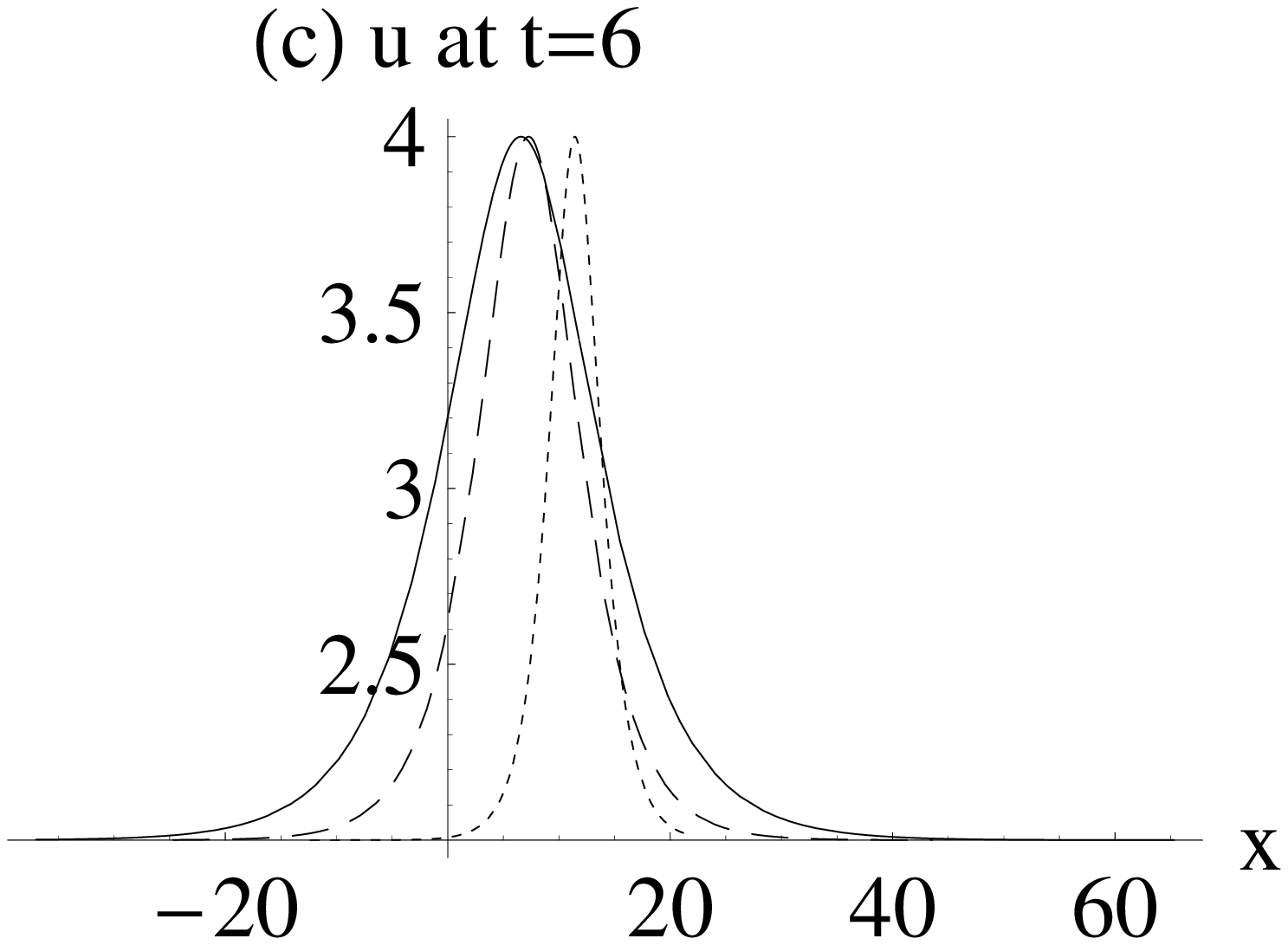}}
\end{center}
\centerline{\small{Figure 2.\,\,One-soliton solution defined by (\ref{36}) with $Q_0=3/2$, $r_0=\sqrt{2}$, $\lambda_1=-1$, }}
\centerline{\small{ the dotted line is for $\alpha=1$, the dashed line is for $\alpha=2$, and the solid line is for $\alpha=3$.}}

\noindent \textbf{Case 2} ($j=2$). According to the Lax pair (\ref{14})-(\ref{15}), their solution can be written as

\begin{equation}\label{37}
\begin{array}{ll}
\overline{\Phi}=T\Phi|_{\lambda=\lambda_2}=
\left(\begin{array}{cc}  -\delta_1   &    1   \\[3mm]
\delta^2_1-\lambda_1+\lambda_2   &  -\delta_1
\end{array}
\right)
\left(\begin{array}{cc}  \sinh\xi_2   \\[3mm]
\Omega_2\cosh\xi_2
\end{array}
\right)\\[7mm]
\mbox{} \hskip 0.3cm=
\left(\begin{array}{cc}  -\delta_1 \sinh\xi_2+ \Omega_2 \cosh\xi_2 \\[3mm]
(\delta^2_1-\lambda_1+\lambda_2) \sinh\xi_2-\delta_1 \Omega_2\cosh\xi_2
\end{array}
\right),
\end{array}
\end{equation}
where $\xi_2$ and $\delta_1$ are given in (\ref{33}) and (\ref{34}), respectively.

Noticing equation (\ref{20}) and (\ref{37}), we have
\begin{equation}\label{38}
\begin{array}{ll}
\delta_2=\dfrac{(-\delta_1 \sinh\xi_2+ \Omega_2 \cosh\xi_2)_y}{-\delta_1 \sinh\xi_2+ \Omega_2 \cosh\xi_2}
=\dfrac{(-\Omega_1\tanh\xi_1\sinh\xi_2+ \Omega_2 \cosh\xi_2)_y}{-\Omega_1\tanh\xi_1 \sinh\xi_2+ \Omega_2 \cosh\xi_2}\\[5mm]
\mbox{} \hskip 0.3cm=\left[\ln (-\Omega_1\tanh\xi_1\sinh\xi_2+ \Omega_2 \cosh\xi_2)\right]_y.
\end{array}
\end{equation}

Based on the formula (\ref{26}), we have
\begin{equation}\label{39}
\begin{array}{ll}
Q_2=Q_0-2\Omega^2_1\sech^2\xi_1-2\left[\ln (-\Omega_1\tanh\xi_1\sinh\xi_2+ \Omega_2 \cosh\xi_2)\right]_{yy},\\[5mm]
r_2=
r_0-\dfrac{r_0\Omega^2_1}{\lambda_1}\sech^2\xi_1-2\alpha^2\left[\ln (-\Omega_1\tanh\xi_1\sinh\xi_2+ \Omega_2 \cosh\xi_2)\right]_{y\tau}.\\[5mm]
\end{array}
\end{equation}
From the formulas (\ref{10}) and (\ref{104}), we arrive at the two-soliton solution of the DGH equation (\ref{2}) in parametric form
\begin{equation}\label{40}
\begin{array}{ll}
u=r^2_2-\alpha^2r_2(\ln r_2)_{y\tau},\\[3mm]
x=\alpha \ln \dfrac{(-\delta_2 \phi^{-}_0+\phi^{-}_{0y})^2}{r_2},\\[3mm]
t=\tau,
\end{array}
\end{equation}
where $\delta_2$ and $r_2$ are given in (\ref{38}) and (\ref{39}), respectively. The interaction of the two-soliton solution (\ref{40}) is shown in figure 3.

\noindent \textbf{Case N} ($j=N$). Generally, we can get $N$-soliton solution of the DGH equation (\ref{2}) as follows
\begin{equation}\label{41}
u=\left(r_0-2\alpha^2\sum^{N}_{j=1}\delta_{j\tau}\right)^2-2\alpha^2\left(r_0-2\alpha^2\sum^{N}_{j=1}\delta_{j\tau}\right)\left(\ln \left(r_0-2\alpha^2\sum^{N}_{j=1}\delta_{j\tau}\right)\right)_{y\tau},
\end{equation}
\begin{equation}\label{42}
\begin{array}{ll}
x=\begin{cases}
\alpha \ln \dfrac{\left(-\delta_N \phi^{+}_0+\phi^{+}_{0y}\right)^2}{r_0-2\alpha^2\sum^{N}_{j=1}\delta_{j\tau}}, & \text{ if $N$ is odd number},\\[5mm]
\alpha \ln \dfrac{\left(-\delta_N \phi^{-}_0+\phi^{-}_{0y}\right)^2}{r_0-2\alpha^2\sum^{N}_{j=1}\delta_{j\tau}}, & \text{ if $N$ is even number},
\end{cases}\\[3mm]
t=\tau,
\end{array}
\end{equation}
where
\begin{equation}\label{43}
\delta_j=\dfrac{(-\delta_{j-1}\phi_j+\phi_{jy})_y}{-\delta_{j-1}\phi_j+\phi_{jy}},\,\,\,
\delta_1=\Omega_1\tanh\xi_1,\,\,\,(j=2,3,\ldots,N),
\end{equation}
and $\phi_j$, $(j=2,3,\ldots,N)$ are given in (\ref{31})-(\ref{32}).
\begin{center}
\resizebox{2.7in}{!}{\includegraphics{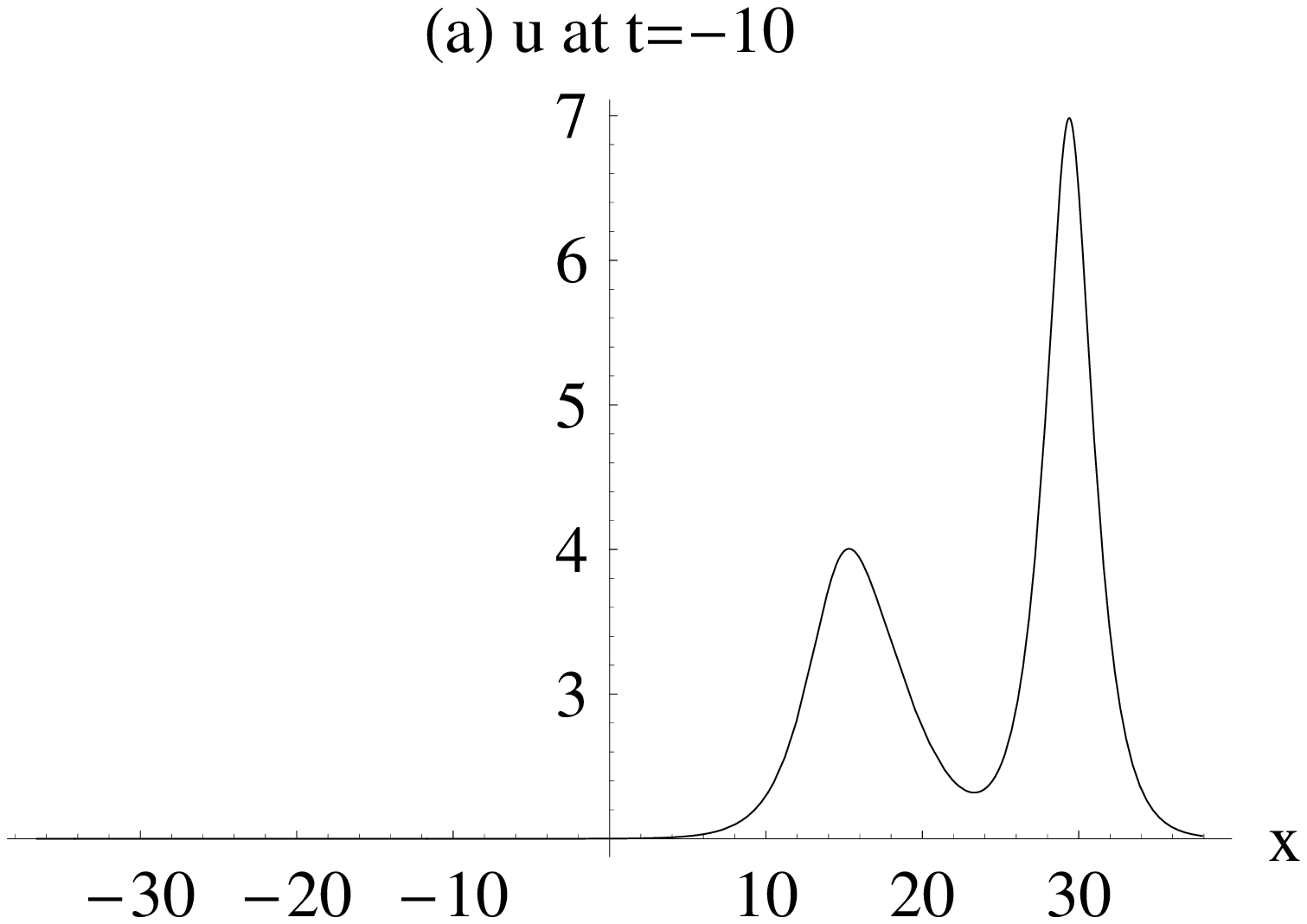}}\,\,\,
\resizebox{2.7in}{!}{\includegraphics{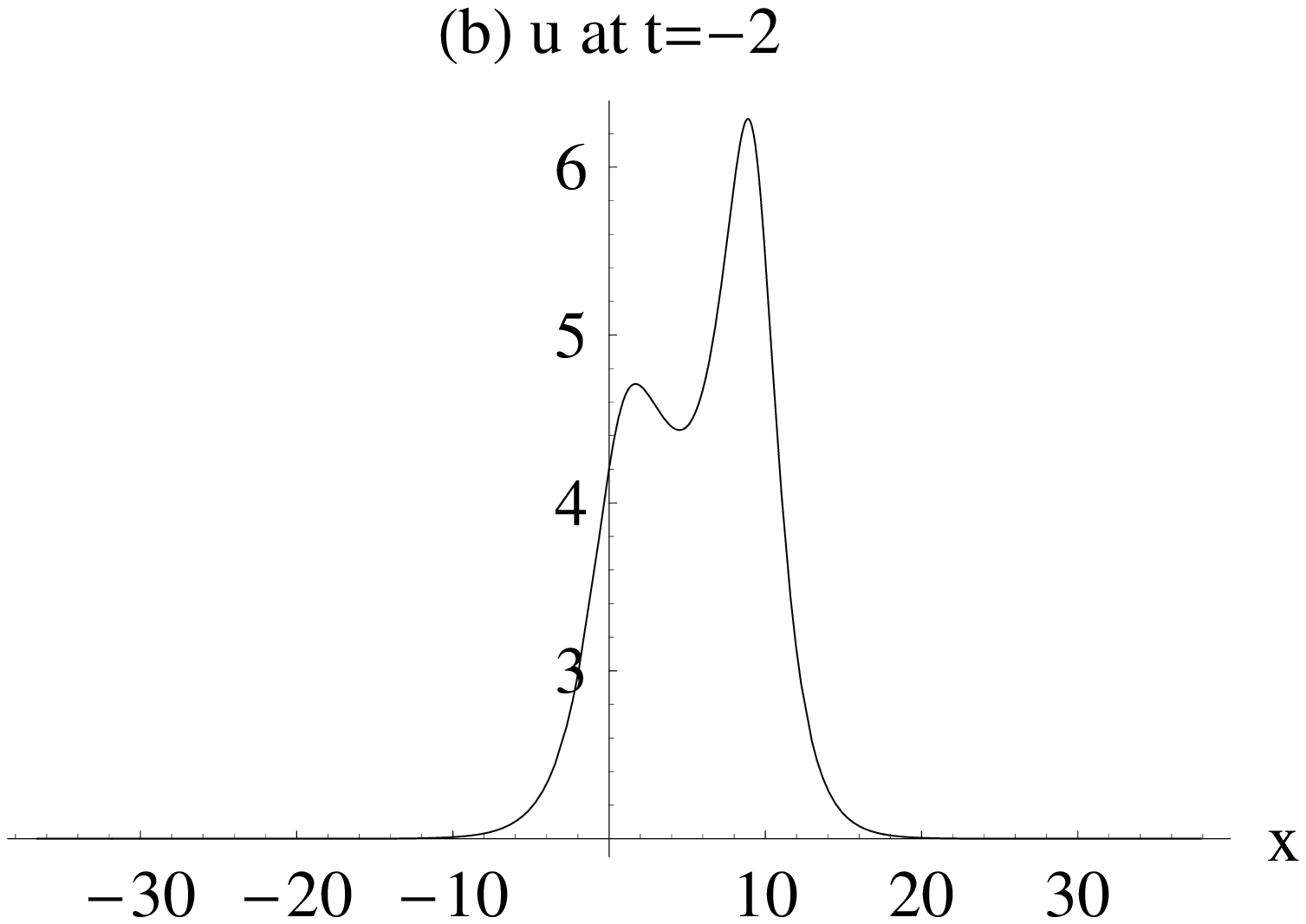}}\,\,\,
\resizebox{2.7in}{!}{\includegraphics{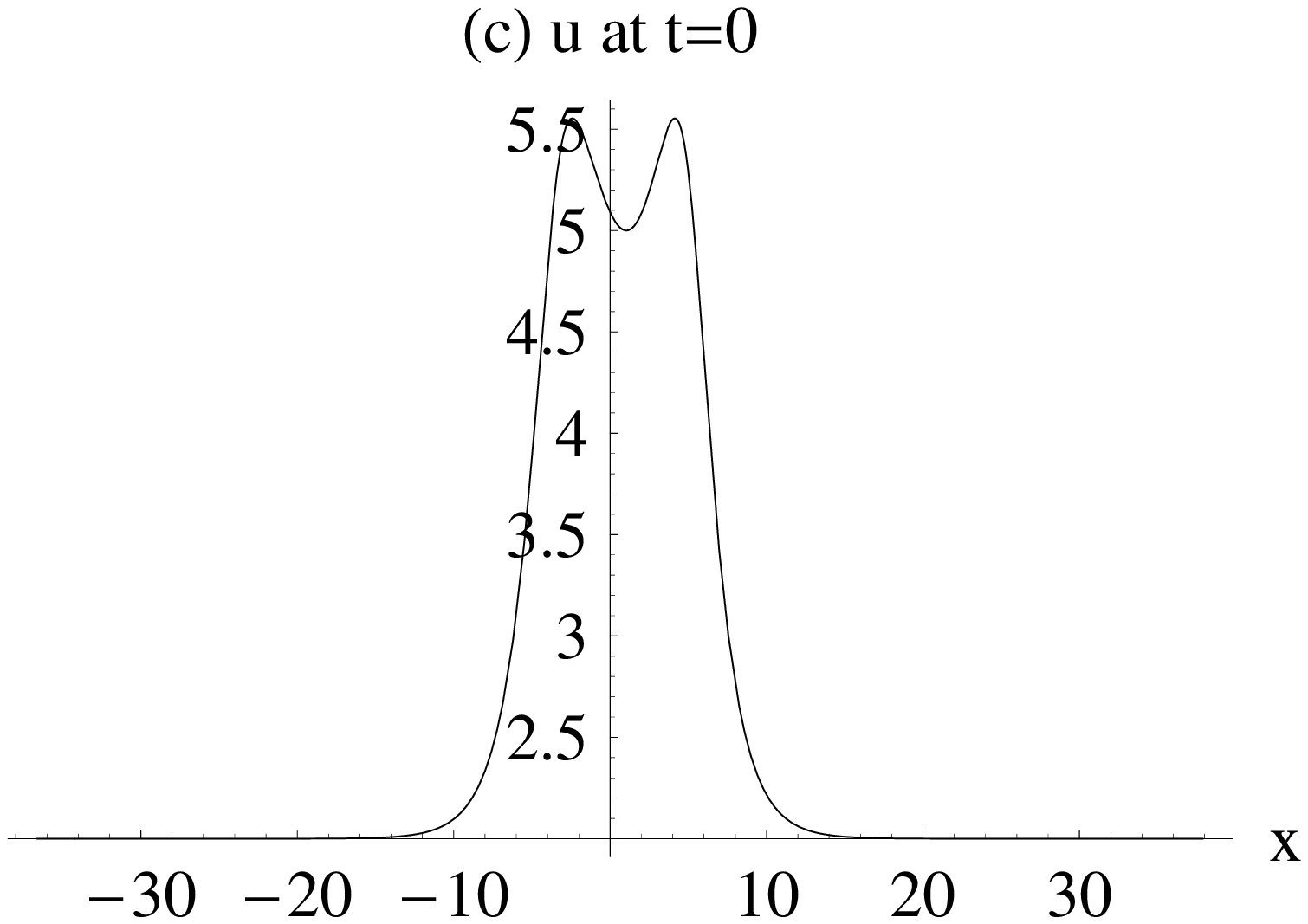}}\,\,\,
\resizebox{2.7in}{!}{\includegraphics{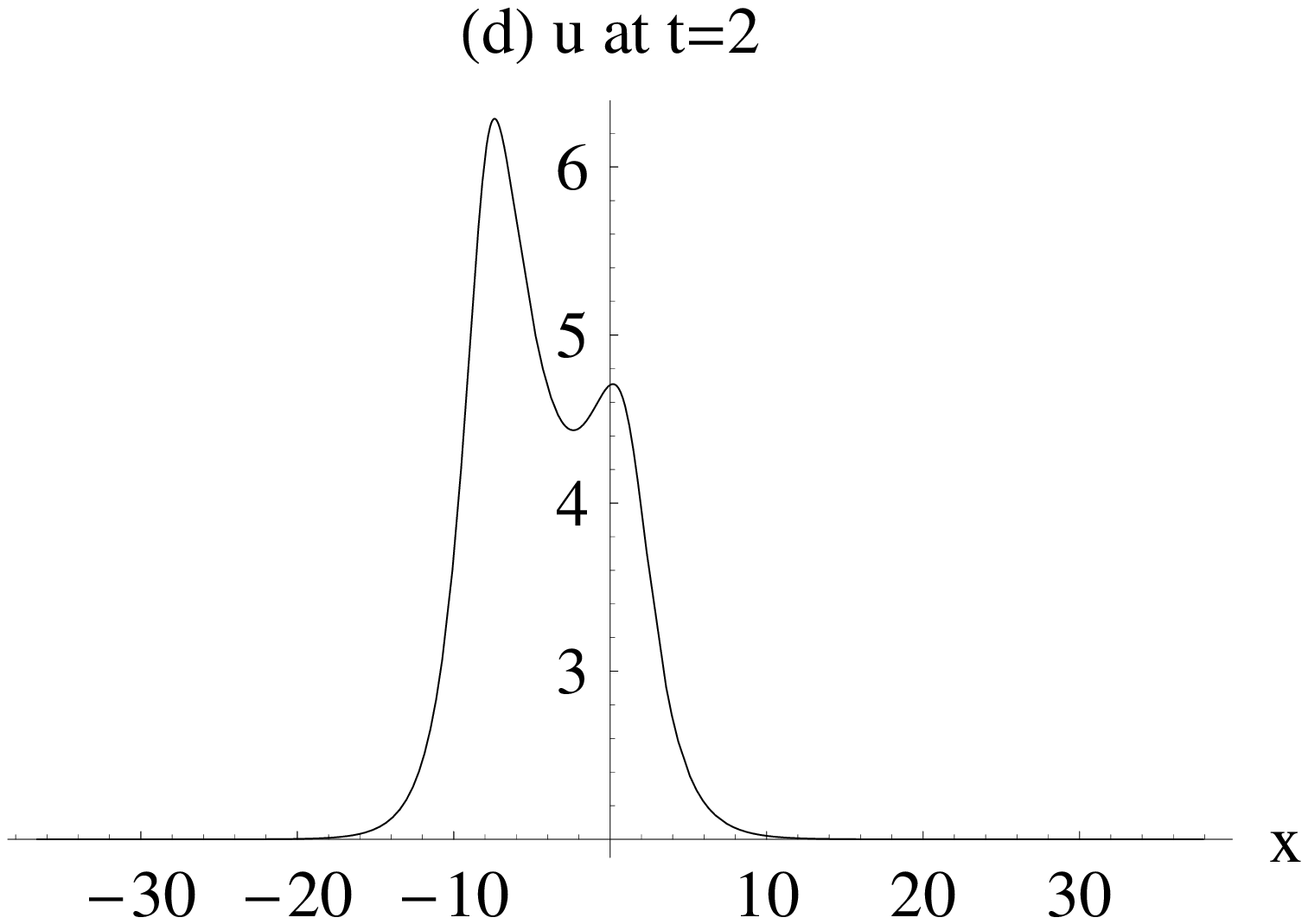}}\,\,\,
\resizebox{2.7in}{!}{\includegraphics{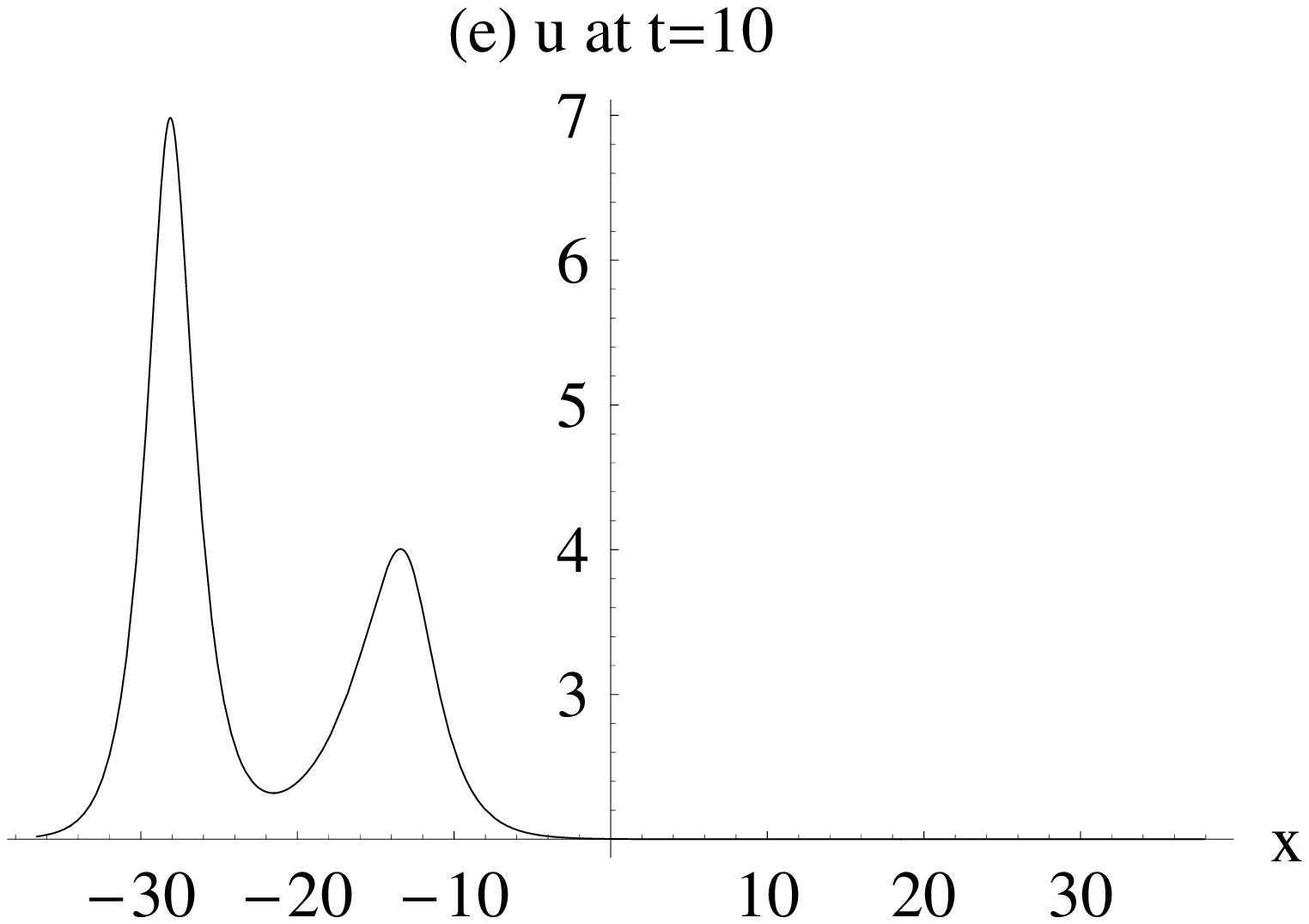}}
\end{center}
\centerline{\small{Figure 3.\,\,Two-soliton solution defined by (\ref{40}) with $Q_0=3/2$, $r_0=\sqrt{2}$, $\lambda_1=-1$, $\lambda_2=-2/3$, $\alpha=1$,}}
\centerline{\small{and (a) at $t=-10$,  (b) at $t=-2$, (c) at $t=0$ , (d) at $t=2$ , (e) at $t=10$.}}

\section{Multi-peakon solution}

In the next, we shall derive multi-peakon type solutions to the DGH equation (\ref{2}).

\noindent \textbf{Case 1} ($n=1$). Let us suppose that one-peakon solution of the DGH equation (\ref{2}) is of
 the following form
 \begin{equation}\label{44}
u(x,t)=p_1(t){\rm e}^{-\left|\xi_1\right|},\,\,\,\xi_1=\dfrac{1}{\alpha}(x-q_{1})
\end{equation}
where $p_{1}(t)$ and $q_{1}(t)$ are functions of $t$
needed to be determined. In this case, it is obvious to see that function $u(x,t)$ do not has the first order derivative at the point $x=q_1(t)$, but we can obtain their derivatives $u_x$, $m$, $m_x$ and $m_t$ in the weak sense as follows
\begin{equation}\label{45}
u_{x}=-\dfrac{p_{1}}{\alpha}sgn\left(\xi_1\right){\rm
e}^{-\left|\xi_1\right|},\,\,\,m=2p_{1}\delta\left(\xi_1\right),
\end{equation}
\begin{equation}\label{46}
m_x=\dfrac{2p_{1}}{\alpha}\delta^\prime\left(\xi_1\right),\,\,\,\,
m_t=2p_{1t}\delta\left(\xi_1\right)-\dfrac{2p_1q_{1t}}{\alpha}\delta^\prime\left(\xi_1\right),
\end{equation}
where  $\delta\left(\xi_1\right)$ denotes delta distribution function.

Substituting (\ref{44})-(\ref{46}) into Eq. (\ref{2}) and integrating in the distribution sense, we can readily get
\begin{equation}\label{47}
p_{1t}=0,\,\,\,q_{1t}=c_0+p_1.
\end{equation}
From (\ref{47}), it is easy to see that we may have
\begin{equation}\label{48}
p_{1}=c,\,\,\,q_{1}=(c_0+c)t+\eta_0,
\end{equation}
where $c$, $c_{0}$  and $\eta_0$ are three arbitrary constants. Substituting
(\ref{48}) into Eq. (\ref{44}), we obtain a one-peakon
solution of the DGH equation (\ref{2}) as follows

\begin{equation}\label{49}
u=c\, {\rm
e}^{-\left|\dfrac{1}{\alpha}(x-\left(c_0+c\right)t-\eta_0)\right|}.
\end{equation}
This one-peakon solution (\ref{49}) was found in \cite{5} by bifurcation method.

\noindent \textbf{Case 2} ($n=2$). We assume that the DGH equation (\ref{2}) admits two-peakon solution as
follows
\begin{equation}\label{50}
u=p_{1}{\rm e}^{-\left|\xi_1\right|}+p_{2}{\rm
e}^{-\left|\xi_2\right|},\,\,\,\,\xi_i=\dfrac{1}{\alpha}(x-q_i),\,\,\,(i=1,2),
\end{equation}
where $p_{1}$ , $p_{2}$ , $q_{1}$  and $q_{2}$
are functions of $t$ needed to be determined. By a direct
calculation in the distribution sense, we have
\begin{equation}\label{51}
u_{x}=-\dfrac{p_{1}}{\alpha}sgn\left(\xi_1\right) {\rm e}^{-\left|\xi_1\right|}-\dfrac{p_{2}}{\alpha}sgn\left(\xi_2\right) {\rm
e}^{-\left|\xi_2\right|},
\end{equation}
\begin{equation}\label{52}
m=2p_{1}\delta
\left(\xi_1\right)+2p_{2}\delta\left(\xi_2\right),\,m_x=\dfrac{2p_{1}}{\alpha}\delta^\prime
\left(\xi_1\right)+\dfrac{2p_{2}}{\alpha}\delta^\prime\left(\xi_2\right),
\end{equation}
\begin{equation}\label{53}
m_t=2p_{1t}\delta\left(\xi_1\right)-\dfrac{2p_{1}q_{1t}}{\alpha}\delta^\prime
\left(\xi_1\right)+2p_{2t}\delta\left(\xi_2\right)-\dfrac{2p_{2}q_{2t}}{\alpha}\delta^\prime\left(\xi_2\right).
\end{equation}
Substituting (\ref{50})-(\ref{53}) into Eq. (\ref{2}) and integrating through test functions yield the following ODE dynamical system

\begin{equation}\label{54}
p_{1t}=\dfrac{2\alpha-1}{\alpha^2}p_1p_2sgn\left(\dfrac{1}{\alpha}(q_{1}-q_{2})\right){\rm e}^{-\left|\dfrac{1}{\alpha}(q_{1}-q_{2})\right|},\,q_{1t}=c_0+p_1+p_2{\rm e}^{-\left|\dfrac{1}{\alpha}(q_{1}-q_{2})\right|},
\end{equation}
\begin{equation}\label{55}
p_{2t}=\dfrac{2\alpha-1}{\alpha^2}p_1p_2sgn\left(\dfrac{1}{\alpha}(q_{2}-q_{1})\right){\rm e}^{-\left|\dfrac{1}{\alpha}(q_{2}-q_{1})\right|},\,q_{2t}=c_0+p_2+p_1{\rm e}^{-\left|\dfrac{1}{\alpha}(q_{2}-q_{1})\right|}.
\end{equation}
We shall discuss all possible values of $\alpha$ in the following cases.

\noindent \textbf{Subcase 2.1} Take $\alpha= \frac{1}{2}$, Eqs. (\ref{54})-(\ref{55}) become the following equations
\begin{equation}\label{56}
p_{1t}=p_{2t}=0,
\,q_{1t}=c_0+p_1+p_2{\rm e}^{-\left|2(q_{1}-q_{2})\right|},\,
\,q_{2t}=c_0+p_2+p_1{\rm e}^{-\left|2(q_{2}-q_{1})\right|}.
\end{equation}
Solving Eq. (\ref{56}), we have
\begin{equation}\label{57}
p_{1}=c_1,\,q_{1}=(c_1+c_0)t-\dfrac{c_2}{c_1-c_2}\ln(1+{\rm e}^{-(c_1-c_2)t-\eta_0})+C_1,
\end{equation}
\begin{equation}\label{58}
p_{2}=c_2,\,q_{2}=(c_2+c_0)t-\dfrac{c_1}{c_1-c_2}\ln(1+{\rm e}^{-(c_1-c_2)t-\eta_0})+C_2,
\end{equation}
where $c_0$, $c_1$, $c_2$, $C_1$, $C_2$, and $\eta_0$ are constants. Substituting (\ref{57})-(\ref{58}) into (\ref{50}), we get a two-peakon solution
\begin{equation}\label{59}
u=c_{1}{\rm e}^{-\left|2(x-q_{1})\right|}+c_{2}{\rm
e}^{-\left|2(x-q_{2})\right|},
\end{equation}
where $q_1$ and $q_2$ are given in (\ref{57})-(\ref{58}). The dynamic behaviors of the two-peakon wave (\ref{59}) are shown in figure 4.
\begin{center}
\resizebox{2.9in}{!}{\includegraphics{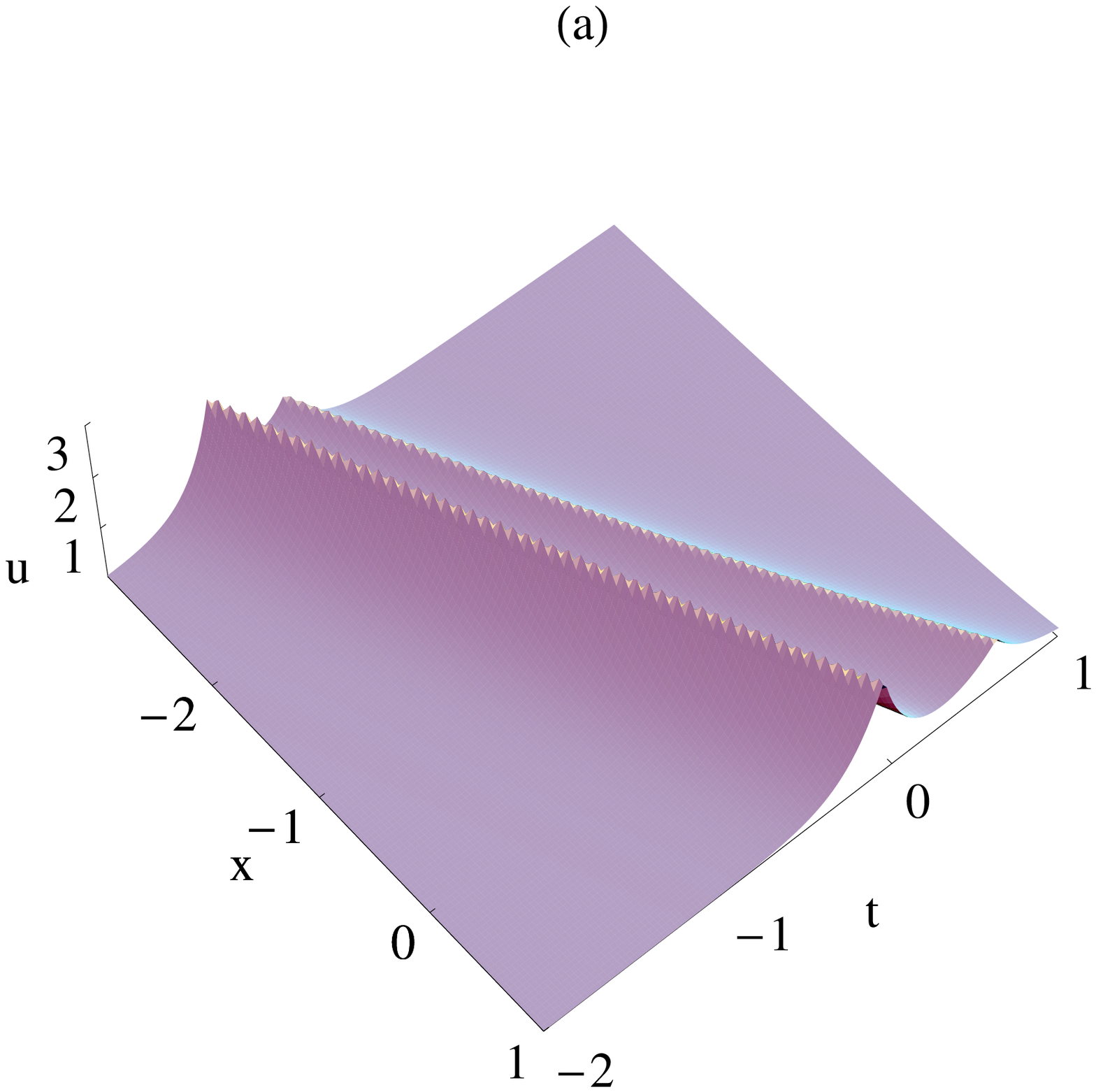}}
\end{center}
\centerline{\small{Figure 4.\,\,Two-peakon solution defined by (\ref{59}) with $c_0=\frac{1}{2}$, $c_1=2$, $c_2=1$, $C_1=2$, $C_2=1$, $\eta_0=0$.}}

\noindent \textbf{Subcase 2.2} Take $\alpha\neq \frac{1}{2}$, $\alpha\neq 0$  and $\Lambda=\frac{2\alpha-1}{\alpha^2}$ in (\ref{54})-(\ref{55}), we have
\begin{equation}\label{60}
(p_1+p_2)_t=0.
\end{equation}
By the above equation, it is easy to see that there are the
following relations
\begin{equation}\label{61}
p_{1}=p,\,\,\,p_{2}=-p+\Gamma,
\end{equation}
where $p=p(t)$ is a function of $t$, and  $\Gamma$ is a integration constant.

Without loss of generality, we take $Q=\frac{1}{\alpha}(q_{1}-q_{2})>0$ and combine Eqs. (\ref{54})-(\ref{55}), we are able to get
\begin{equation}\label{62}
\alpha Q_{t}=\left(2p-\Gamma\right)\left(1-{\rm e}^{-Q}\right).
\end{equation}
From the first equation of (\ref{54})-(\ref{55}) and (\ref{61}), we have
\begin{equation}\label{63}
Q=-\ln\left(-\dfrac{p_t}{\Lambda p(p-\Gamma)}\right).
\end{equation}
Combining (\ref{62}) and (\ref{63}) leads to
\begin{equation}\label{64}
\alpha \Lambda p(p-\Gamma)p_{tt}+\Lambda p(\Gamma-p)(\Gamma-2p)p_t+(\alpha \Lambda-1)(\Gamma-2p)p^2_t=0.
\end{equation}

For Eq. (\ref{64}), we shall give some examples as the follows.

\noindent \textbf{Example 1.} Let choosing $\alpha=1$, we have
\begin{equation}\label{65}
p_{t}+p^2-\Gamma p=d,
\end{equation}
where $d$ is a integrate constant. Based on  the solutions of the Eq.
(\ref{65}), we shall construct three type of two-peakon solutions of the DGH equation (\ref{2}).

$\mathbf{(i)}$ For $d>-\dfrac{\Gamma^{2}}{4}$, Eq.
(\ref{65}) leads to
\begin{equation}\label{66}
p=\dfrac{\Gamma}{2}+\Delta\coth\Delta(t-\eta_0),\,\,\,\Delta=\sqrt{d+\dfrac{1}{4}\Gamma^2}.
\end{equation}

Substituting (\ref{63}) and (\ref{66}) into the second equation of (\ref{54})-(\ref{55}), we obtain
\begin{equation}\label{67}
q_1=\left(\dfrac{\Gamma}{2}+c_0\right)t+\ln\left|\dfrac{\Gamma}{2}\sinh\Delta(t-\eta_0)+\Delta\cosh\Delta(t-\eta_0)\right|+C_1,
\end{equation}
\begin{equation}\label{68}
q_2=\left(\dfrac{\Gamma}{2}+c_0\right)t-\ln\left|-\dfrac{\Gamma}{2}\sinh\Delta(t-\eta_0)+\Delta\cosh\Delta(t-\eta_0)\right|+C_2,
\end{equation}
where $\Delta=\sqrt{d+\dfrac{1}{4}\Gamma^2}$ and $\eta_0$, $C_1$, $C_2$ are integrable constants.

Therefore, we have the first type two-peakon solutions of the DGH equation
(\ref{2}) as
\begin{equation}\label{69}
u=\left[\dfrac{\Gamma}{2}+\Delta\coth\Delta(t-\eta_0)\right]{\rm e}^{-|x-q_1|}+\left[\dfrac{\Gamma}{2}-\Delta\coth\Delta(t-\eta_0)\right]{\rm e}^{-|x-q_2|},
\end{equation}
where $q_1$ and $q_2$ are given in (\ref{67})-(\ref{68}). In figure 5, we have described the interaction processes of the first type two-peakon solutions (\ref{69}).  The head on elastic collision of two-peakon waves (\ref{69}) is illustrated in figure 5. This interaction is very similar to that of two peakons of the CH equation.

$\mathbf{(ii)}$ For $d<-\dfrac{\Gamma^{2}}{4}$, Eq. (\ref{65}) leads to

\begin{equation}\label{70}
p=\dfrac{\Gamma}{2}+\Omega\tan\Omega(t-\eta_0),\,\,\,\Omega=\sqrt{-d-\dfrac{1}{4}\Gamma^2}.
\end{equation}

Substituting (\ref{63}) and (\ref{70}) into the second equation of (\ref{54})-(\ref{55}), we obtain
\begin{equation}\label{71}
q_1=\left(\dfrac{\Gamma}{2}+c_0\right)t-2\ln|\cos\Omega(t-\eta_0)|+\ln\left|\dfrac{\Gamma}{2}\cos\Omega(t-\eta_0)+\Omega\sin\Omega(t-\eta_0)\right|+C_1,
\end{equation}
\begin{equation}\label{72}
q_2=\left(\dfrac{\Gamma}{2}+c_0\right)t+2\ln|\cos\Omega(t-\eta_0)|-\ln\left|-\dfrac{\Gamma}{2}\cos\Omega(t-\eta_0)+\Omega\sin\Omega(t-\eta_0)\right|+C_2,
\end{equation}
where $\Omega=\sqrt{-d-\dfrac{1}{4}\Gamma^2}$ and $\eta_0$, $C_1$, $C_2$ are integrable constants.

So we have second type two-peakon solutions for the DGH equation (\ref{2})
\begin{equation}\label{73}
u=\left[\dfrac{\Gamma}{2}+\Omega\tan\Omega(t-\eta_0)\right]{\rm e}^{-|x-q_1|}+
\left[\dfrac{\Gamma}{2}-\Omega\tan\Omega(t-\eta_0)\right]{\rm e}^{-|x-q_2|},
\end{equation}
where $q_1$ and $q_2$ are given in (\ref{71})-(\ref{72}). In figure 6, the period structure of the second type two-peakon
solutions (\ref{73}) is described. It can be seen that this solution (\ref{73}) has singularities.
\begin{center}
\resizebox{2.9in}{!}{\includegraphics{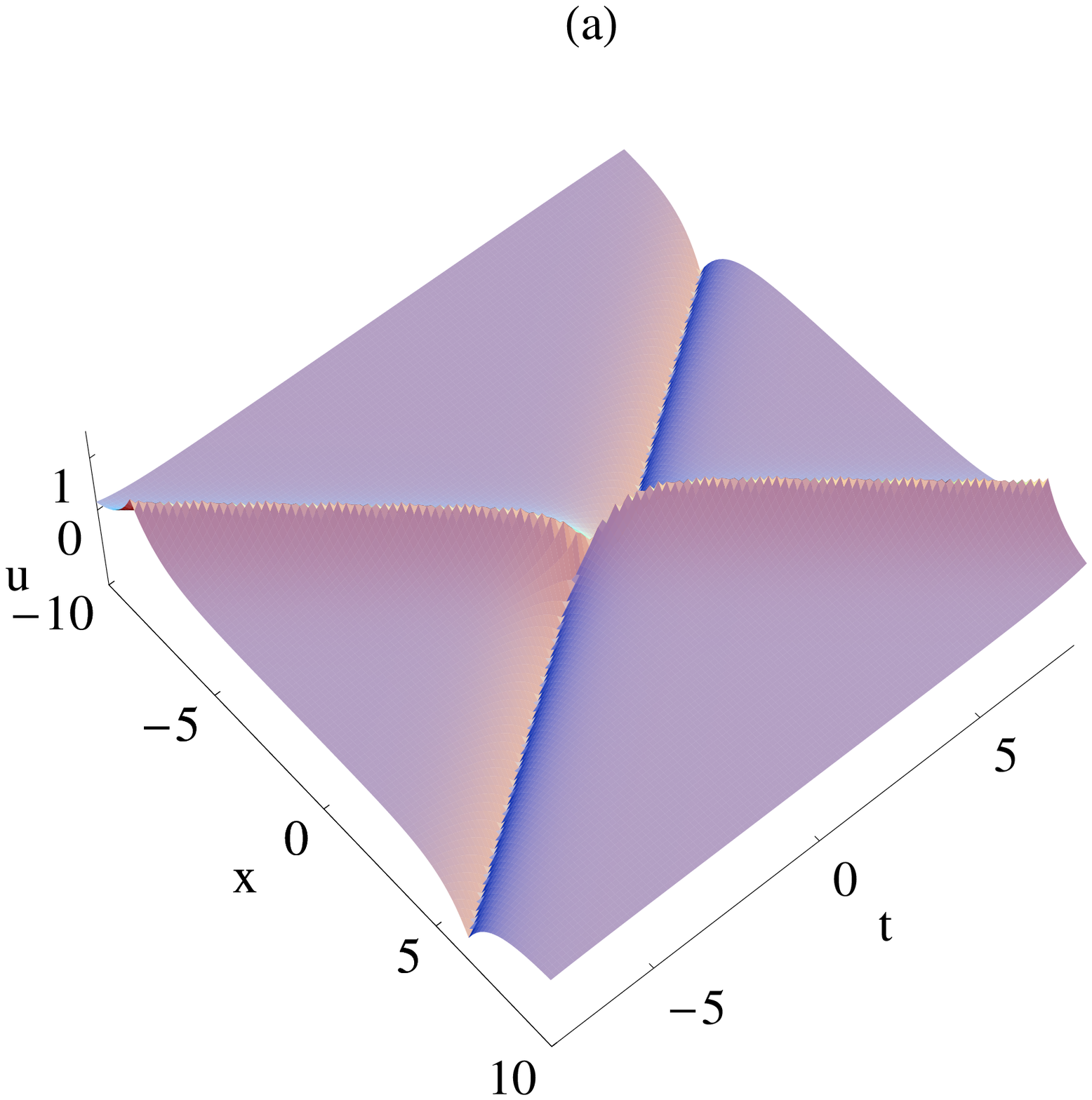}}
\end{center}
\centerline{\small{Figure 5.\,\,The first type two-peakon solution defined by (\ref{69}) with $c_0=\frac{1}{10}$, $d=1$, $C_1=C_2=\eta_0=0$, $\Gamma=\frac{1}{100}$.}}

\begin{center}
\resizebox{2.9in}{!}{\includegraphics{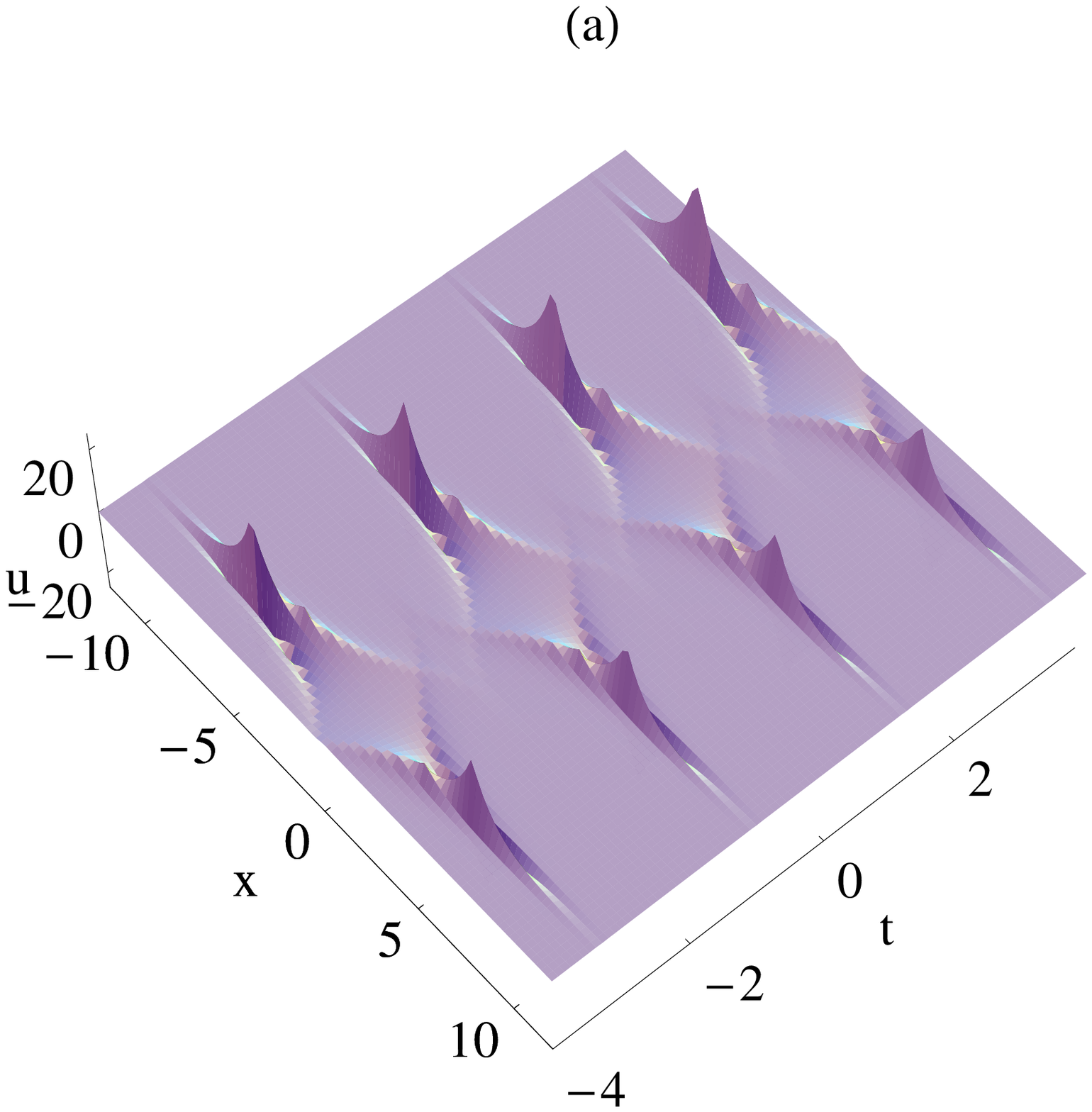}}
\end{center}
\centerline{\small{Figure 6.\,\,The second type two-peakon solution defined by (\ref{73}) with $c_0=-\frac{1}{100}$, $d=-2$, $C_1=C_2=\eta_0=\Gamma=0$.}}

$\mathbf{(iii)}$ For $d=-\dfrac{\Gamma^{2}}{4}$, Eq. (\ref{65}) leads to
\begin{equation}\label{74}
p=\dfrac{\Gamma}{2}+\dfrac{1}{t-\eta_0},
\end{equation}
where $\eta_0$ is an integrable constant.

Substituting (\ref{63}) and (\ref{74}) into the second equation of (\ref{54})-(\ref{55}), we obtain
\begin{equation}\label{75}
q_1=\left(\dfrac{\Gamma}{2}+c_0\right)t+\ln\left|\dfrac{\Gamma}{2}(t-\eta_0)+1\right|+C_1,
\end{equation}
\begin{equation}\label{76}
q_2=\left(\dfrac{\Gamma}{2}+c_0\right)t-\ln\left|-\dfrac{\Gamma}{2}(t-\eta_0)+1\right|+C_2,
\end{equation}
where $\eta_0$, $C_1$ and $C_2$ are integrable constants.

So we have the third type two-peakon solutions for the DGH equation (\ref{2})
\begin{equation}\label{77}
u=\left[\dfrac{\Gamma}{2}+\dfrac{1}{t-\eta_0}\right]{\rm e}^{-|x-q_1|}+\left[\dfrac{\Gamma}{2}-\dfrac{1}{t-\eta_0}\right]{\rm e}^{-|x-q_2|},
\end{equation}
where $q_1$ and $q_2$ are given in (\ref{75})-(\ref{76}). Figure 7 show the profiles of the third type two-peakon
solutions (\ref{77}).
\begin{center}
\resizebox{2.9in}{!}{\includegraphics{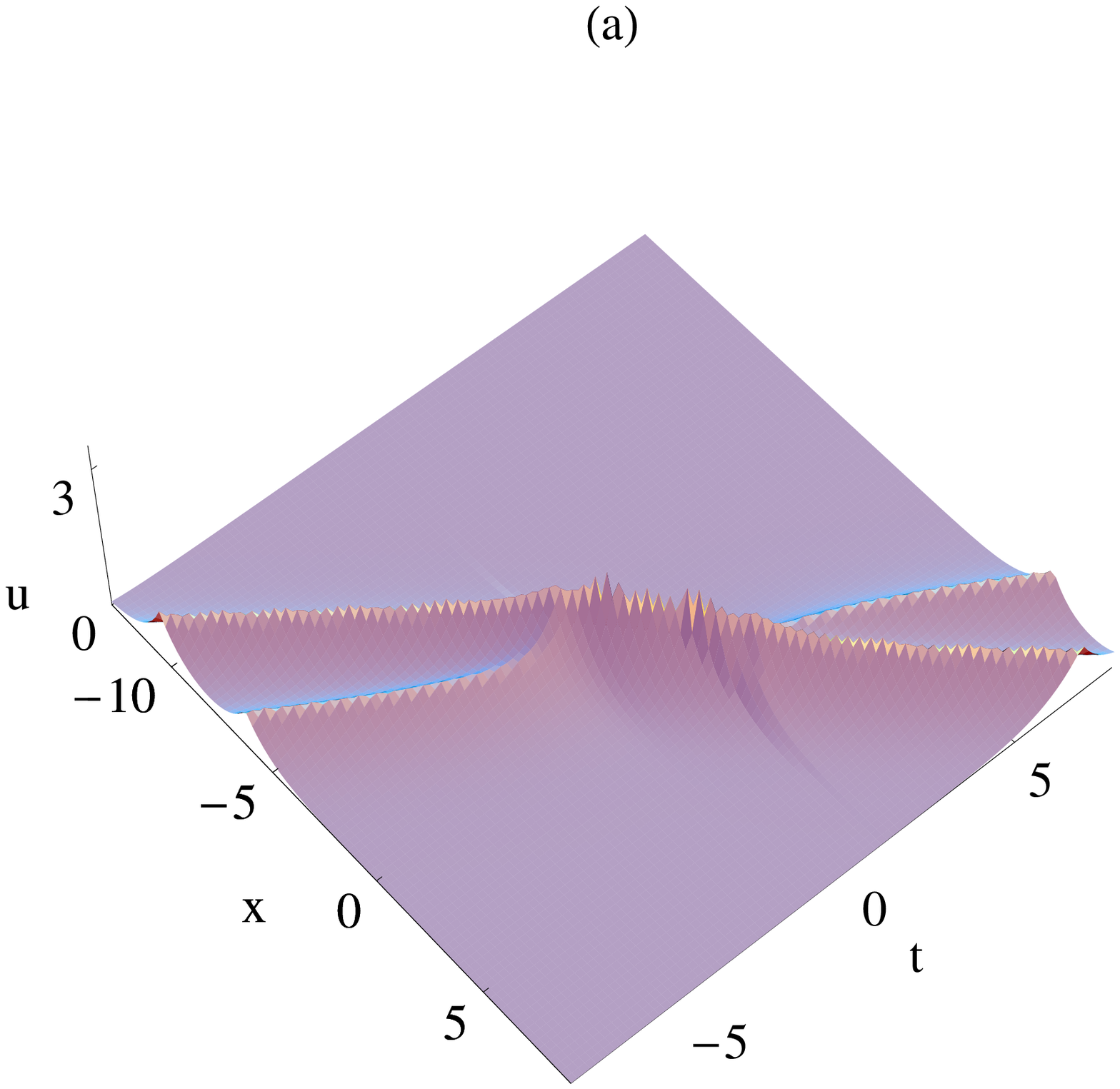}}
\end{center}
\centerline{\small{Figure 7.\,\,The third type two-peakon solution defined by (\ref{77}) with $c_0=\frac{1}{100}$, $C_1=C_2=\eta_0=0$, $\Gamma=2$.}}

\noindent \textbf{Example 2.} For $\Lambda\neq 0$, we obtain a particular solutions of Eq. (\ref{64}) as

\begin{equation}\label{78}
p=\dfrac{A \Gamma}{A+B {\rm e}^{-\Lambda\Gamma t}},
\end{equation}
where $A$ and $B$ are integrable constants. Substituting (\ref{78}) into (\ref{61}) and (\ref{54})-(\ref{55}), we get
\begin{equation}\label{79}
p_1=\dfrac{A \Gamma}{A+B {\rm e}^{-\Lambda\Gamma t}},\,\,p_2=-\dfrac{A \Gamma}{A+B {\rm e}^{-\Lambda\Gamma t}}+\Gamma,
\end{equation}
\begin{equation}\label{80}
q_1=(\Gamma+c_0)t+C_1,\,\,\,q_2=(\Gamma+c_0)t+C_2,
\end{equation}
\begin{center}
\resizebox{2.6in}{!}{\includegraphics{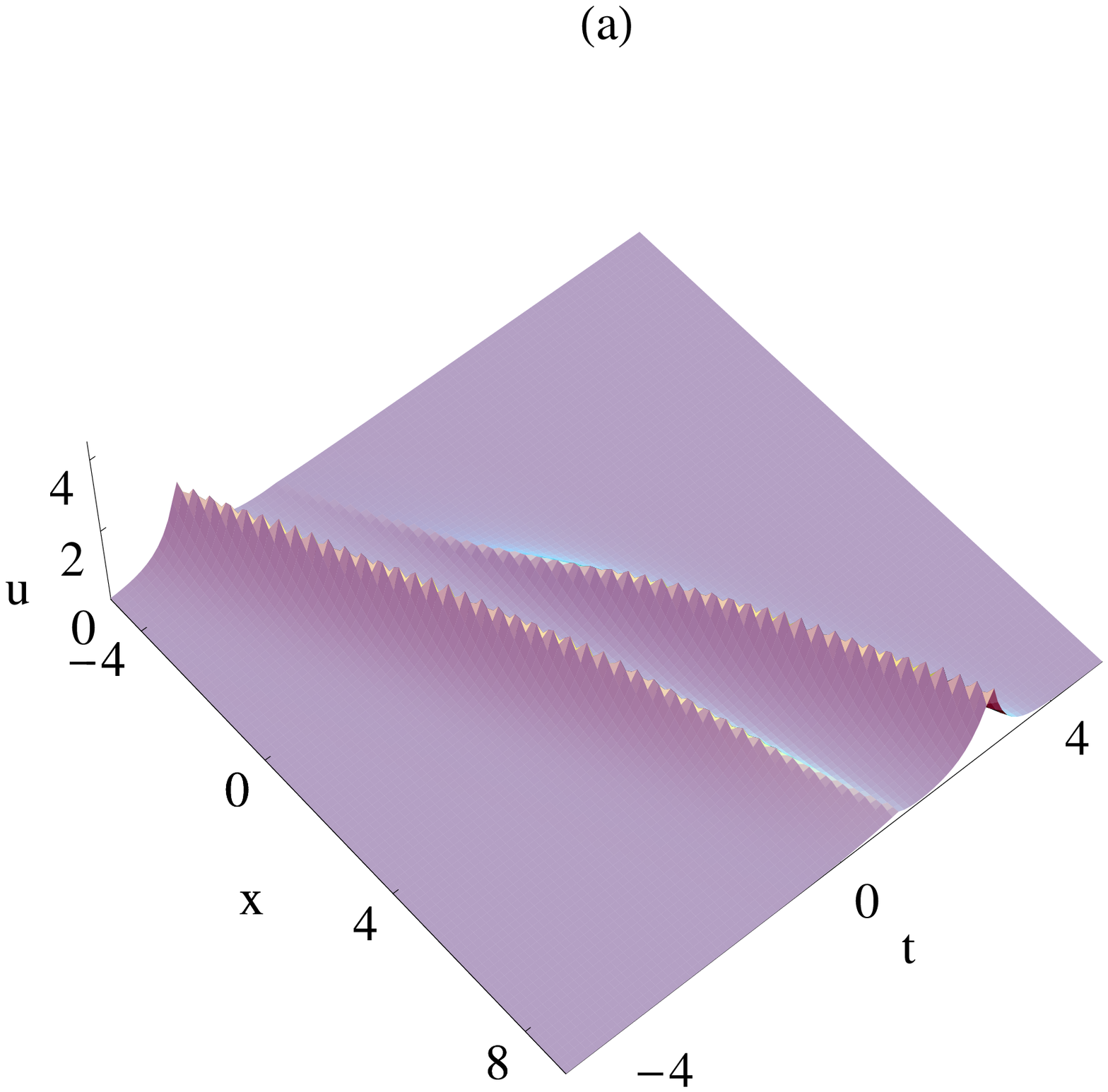}}\,\,\,\,
\resizebox{2.6in}{!}{\includegraphics{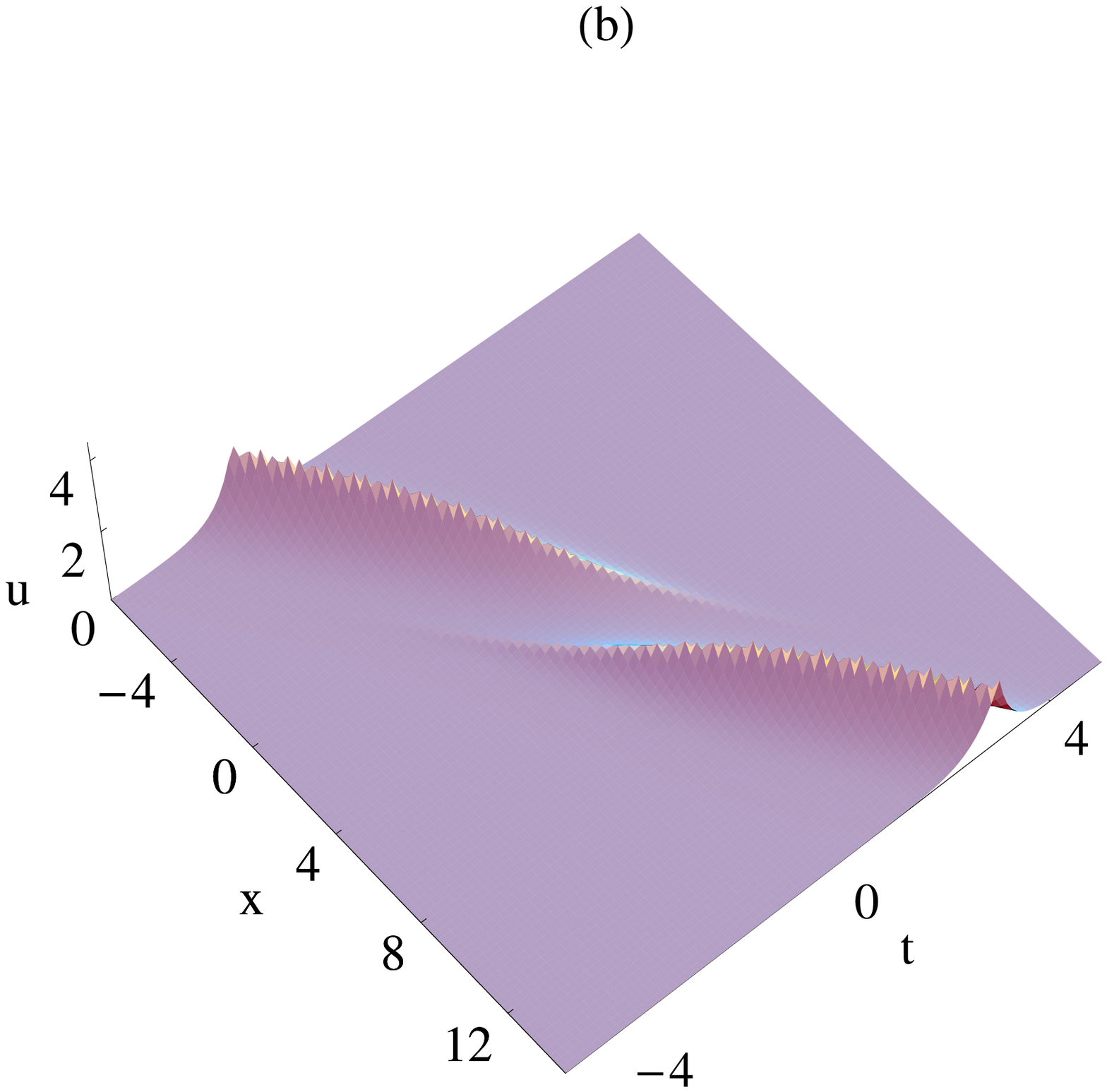}}\,\,\,\,
\resizebox{2.6in}{!}{\includegraphics{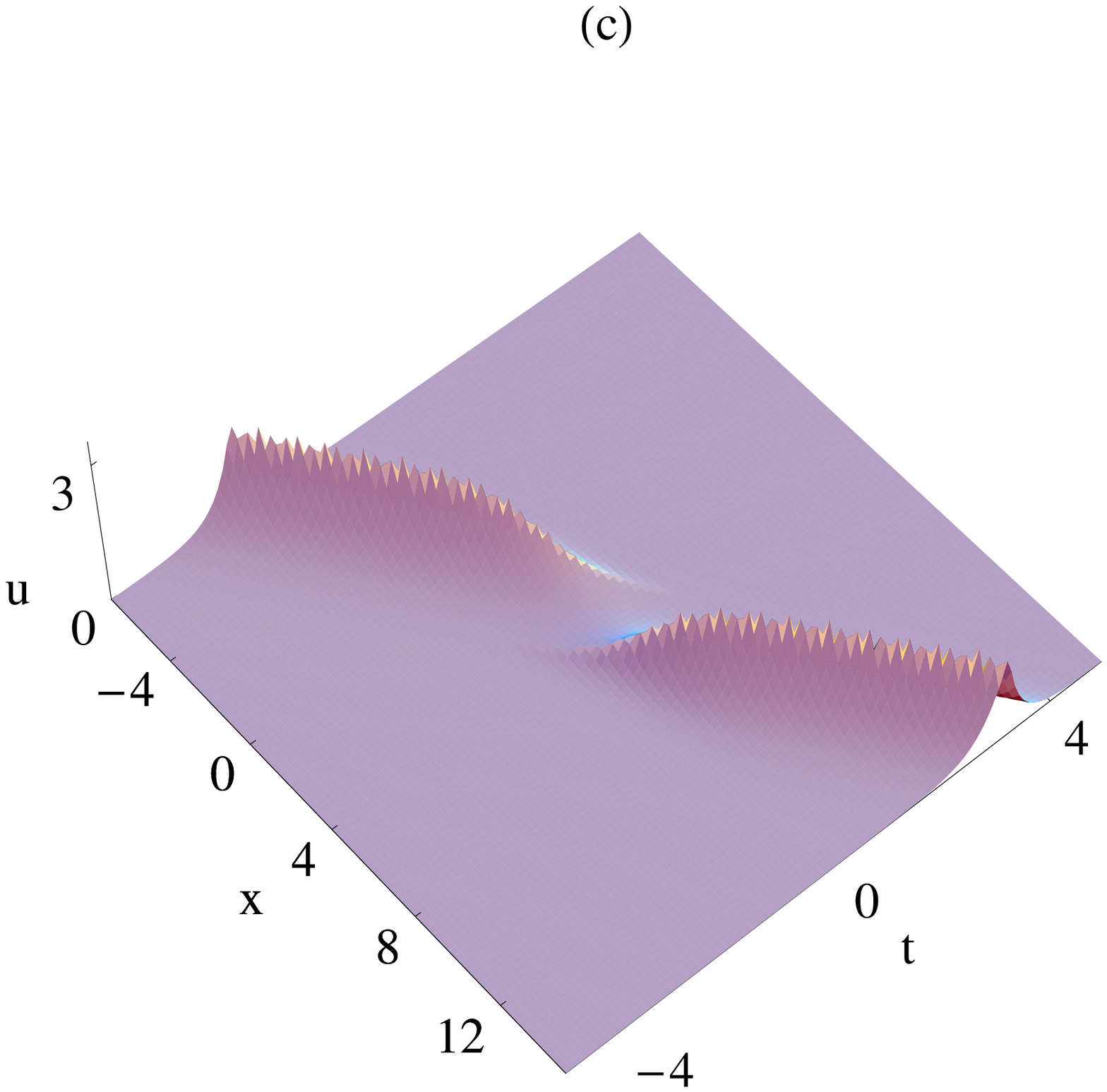}}\,\,\,\,
\resizebox{2.6in}{!}{\includegraphics{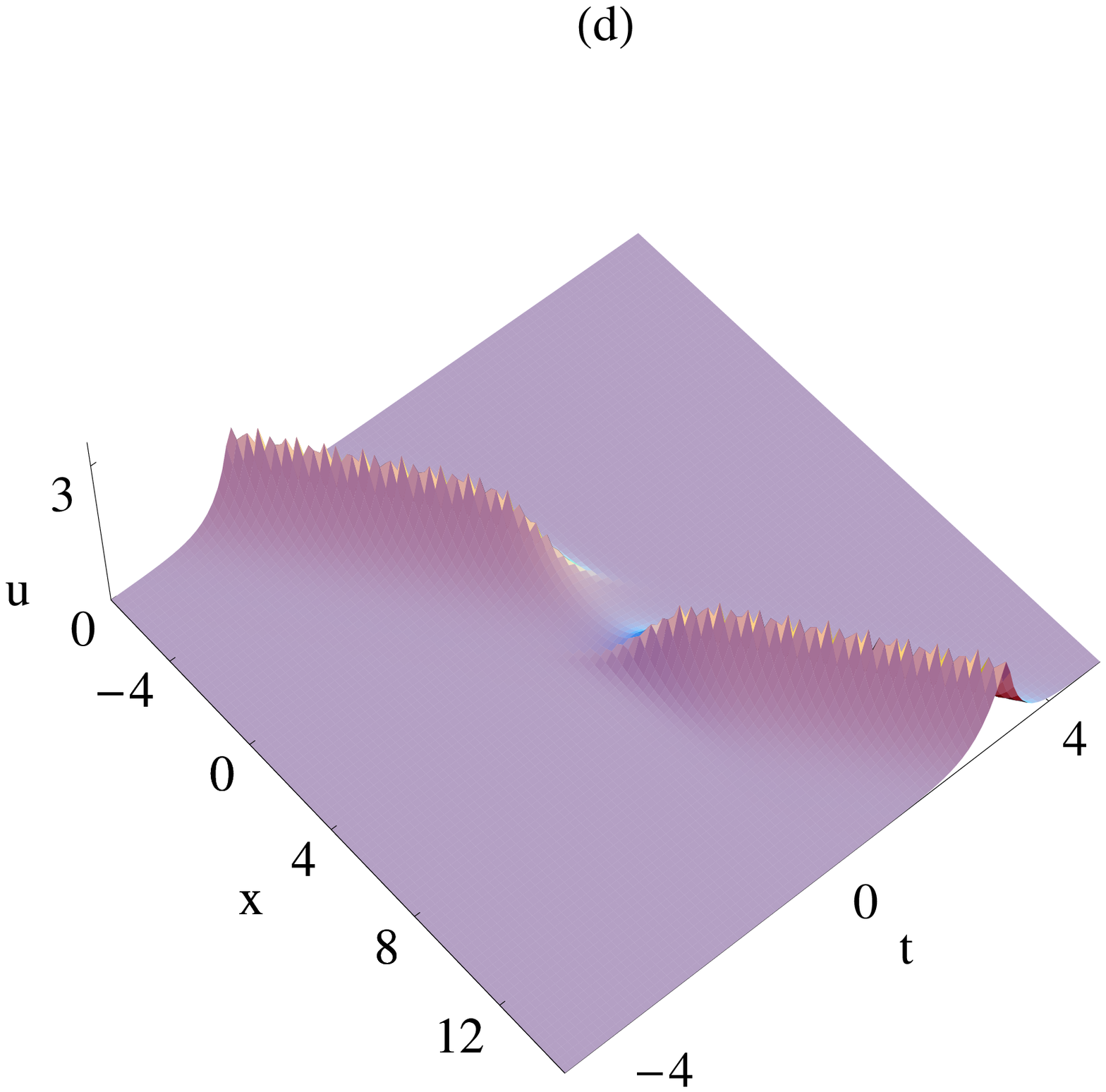}}\,\,\,\,
\resizebox{2.6in}{!}{\includegraphics{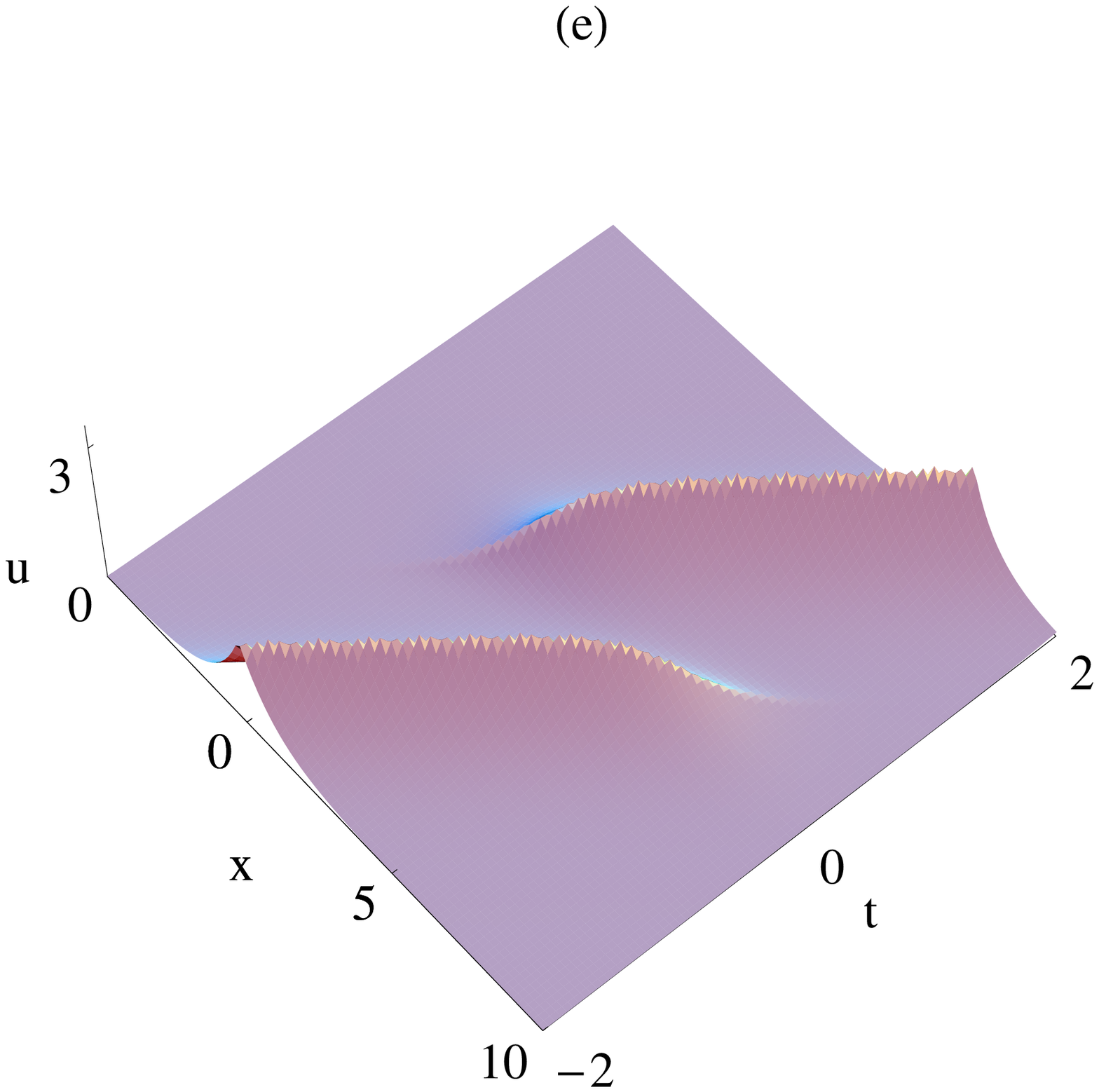}}\,\,\,\,
\end{center}
\centerline{\small{Figure 8.\,\,The solitoff structure in the DGH equation  defined by (\ref{81}) with $C_1=6$, $C_2=1$, $\Gamma=2$, $c_0=A=B=1$,}}
\centerline{\small{ and (a) at $\alpha=-1+\sqrt{2}$,  (b) at $\alpha=1$, (c) at $\alpha=\frac{1}{3}(1+{\rm i}\sqrt{2})$, (d) at $\alpha=\frac{1}{2}(1+{\rm i})$, (e) at $\alpha=\frac{1}{3}$.}}
where $C_1$ and $C_2$ are integrable constants. Substituting (\ref{79})-(\ref{80}) into (\ref{50}), we have a solution of the DGH equation
\begin{equation}\label{81}
u=\dfrac{A \Gamma}{A+B {\rm e}^{-\Lambda\Gamma t}}\, {\rm e}^{-|x-q_1|}+\left[-\dfrac{A \Gamma}{A+B {\rm e}^{-\Lambda\Gamma t}}+\Gamma\right]{\rm e}^{-|x-q_2|},
\end{equation}
where $q_1$ and $q_2$ are given in (\ref{80}).  Figure 8 plot the solitoff structure of the DGH equation for several values of $\alpha$.

\noindent \textbf{Case N} ($n=N$). In general, we suppose $N$-peakon solution of the DGH equations (\ref{2}) has the following form
\begin{equation}\label{82}
u=\sum^n_{j=1}p_j(t){\rm e}^{-|\frac{1}{\alpha}(x-q_j(t))|}.
\end{equation}
Substituting (\ref{82}) into the DGH equation (\ref{2}) and integrating through test functions, we obtain the $N$-peakon dynamical system as follows
\begin{equation}\label{46000}
p_{jt}=\Lambda p_j\sum^N_{k=1}p_k sgn[\frac{1}{\alpha}\left(q_j-q_k\right)]{\rm
e}^{-\left|\frac{1}{\alpha}(q_{j}-q_{k})\right|},
\end{equation}
\begin{equation}\label{47000}
q_{jt}=\sum^N_{k=1}p_k {\rm e}^{-\left|\frac{1}{\alpha}(q_{j}-q_{k})\right|}+c_0,\,\,\,(j=1,2,\ldots,N).
\end{equation}
Thus, $N$-peakon solutions of the DGH equation are obtained by simply superimposing the single peakon solutions and solving for the evoluting of  their amplitudes $p_j$ and the positions of their peakons $q_j$.

\section{Conclusions}

In this paper, the DGH equation (\ref{2}) is mapped to a negative order KdV equation with the help of the reciprocal transformation. And the multi-soliton solutions of the DGH equation (\ref{2}) are obtained by the Darboux transformation to the negative order KdV equation. Further, the multi-peakon solution of the DGH equation (\ref{2}) are obtained by direct computation. And also, the numerical simulations for multi-soliton solutions and multi-peakon solution of the DGH equation (\ref{2}) are given. 

By comparing with known results of the DT \cite{17,18} and direct computation method \cite{15}, our main achievements are as follows:

(i) The part of this paper refines Xia-Zhou-Qiao's procedure \cite{18} for solving the DGH equation (\ref{2}) for a multiple soliton solutions by Darboux matrix approach. Some proofs of the DT and the reciprocal transformation are discussed in detail. The parameter $\alpha$ plays an important role in the soliton solution of the DGH equation (\ref{2}). As $\alpha$ increases with value, the wave becomes more and more widely, as well as its velocity becomes more and more slowly (see figure 2). An interesting phenomenon was found in \cite{17}. Such as the soliton with smaller amplitude can travel faster than the one with larger amplitude when they interact. In this paper, we arrive at the usual result that a soliton with larger amplitude travels faster (see figure 3).

(ii) By classification of the values of $\alpha$,  several kinds of the multi-peakon solutions of the DGH equation (\ref{2}) are obtained in this paper, which including the constant amplitudes, the hyperbolic function amplitudes, the trigonometric function amplitudes, the rational function amplitudes, and the exponential function amplitudes. The peakon-antipeakon interaction in the DGH equation (\ref{2}) was given in \cite{15}, which is very similar to our hyperbolic function result (\ref{69}) (see figure 5). The rest of the multi-peakon solutions of the DGH equation (\ref{2}) are novel.

We believe that these multi-solitons and multi-peakons would help us in recognizing the interaction behaviors of the the DGH equation (\ref{2}).

\section*{Acknowledgments}
This work is supported by the National Natural Science Foundation of
China under (Grant No 11261037), the Natural Science Foundation of
Inner Mongolia Autonomous Region under (Grant No 2014MS0111), the
Caoyuan Yingcai Program of Inner Mongolia Autonomous Region under
(Grant No CYYC2011050), the Program for Young Talents of Science and
Technology in Universities of Inner Mongolia Autonomous Region under
(Grant No NJYT14A04).

\end {document}